\documentclass[11pt,a4paper]{article}

\usepackage[utf8x]{inputenc}
\usepackage[T1]{fontenc}
\usepackage{color}
\usepackage{amssymb}
\usepackage{amsmath,amsfonts}
\usepackage{graphicx}
\usepackage{mathtools}
\usepackage{ragged2e}
\usepackage[normalem]{ulem}

\usepackage{dcolumn}
\usepackage{bm}
\usepackage{graphicx}
\usepackage{verbatim} 
\usepackage[english]{babel}
\usepackage{hyperref}
\hypersetup{
citecolor=red,
colorlinks=true,
filecolor=red,
linkcolor=blue,
linktocpage=true,
urlcolor=blue
} 
\usepackage[titletoc,toc,title]{appendix}
\numberwithin{equation}{section}
\usepackage{cleveref}
\usepackage[margin=2.5cm]{geometry}
\usepackage{cite}
\usepackage{epsfig}
\usepackage{float}
\usepackage{enumitem}
\usepackage[font={footnotesize,it}]{caption}
\usepackage{cases}
\usepackage{amsmath,amssymb,tikz,hyperref,empheq}

\newcommand\trick[1]{}
\newcommand{\be}{\begin{equation}} 
\newcommand{\ee}{\end{equation}}
\newcommand{\eq}[1]{(\ref{#1})}
\newcommand{\bit}{\begin{itemize}}  \newcommand{\eit}{\end{itemize}}
\newcommand{\ben}{\begin{enumerate}}  \newcommand{\een}{\end{enumerate}}

\def\bd{\begin{document}}
\def\ed{\end{document}}
\def\bea{\begin{eqnarray}}
\def\eea{\end{eqnarray}}
\let\bm=\bibitem

\def\la{\langle}
\def\ra{\rangle}

\def\JY#1{{\color{red}{\bf #1}}}

\newcommand{\rrr}[1]{\vskip 0.2cm \noindent{\bf #1} ---}

\long\def\symbolfootnote[#1]#2{\begingroup%
\def\thefootnote{\fnsymbol{footnote}}\footnote[#1]{#2}\endgroup}
\long\def\RemarkBox#1{\begin{flushleft}\fbox{\begin{minipage}
{17.5cm}{\bf Remark:} ~#1\end{minipage}}\end{flushleft}}

\newcommand{\khu}{{\it Department of Physics, College of Science, Kyung Hee University, 26, Kyunghee-daero, Dongdaemun-gu, Seoul 02447, Republic of Korea}}
\newcommand{\ribs}{{\it Research Institute for Basic Sciences, Kyung Hee University, 26, Kyunghee-daero, Dongdaemun-gu, Seoul 02447, Republic of Korea}}
\newcommand{\iriqm}{{\it International Research Institute for Quantum Materials, Kyung Hee University, 26, Kyunghee-daero, Dongdaemun-gu, Seoul 02447, Republic of Korea}}

\begin{document}
\begin{center}
~\vspace{20pt}
  
\thispagestyle{empty}
              {\Large \bf Timelike entanglement entropy with gravitational anomaly from complexified geometry}
\vspace{30pt}

 Minwoo Lim${}^{1}$\symbolfootnote[1]
 {Email:~\tt mim55449@khu.ac.kr},
 Himanshu Parihar${}^{1,2}$\symbolfootnote[2]
 {Email:~\tt himansp@khu.ac.kr},
 Junggi Yoon${}^{1,2,3}$\symbolfootnote[3]
 {Email:~\tt junggi.yoon@khu.ac.kr}

\vspace{0.4cm}

\vspace{5pt}${{}^{1}}$\khu

\vspace{5pt}${{}^{2}}$\ribs

\vspace{5pt}${{}^{3}}$\iriqm

\vspace{5mm}

\vspace{1cm}

\begin{abstract}

We investigate the timelike entanglement entropy (TEE) in a two-dimensional conformal field theory with a gravitational anomaly, $c_L\neq c_R$, dual to topologically massive gravity in AdS$_3$. Holographically, the TEE is determined by the on-shell action of a massive spinning particle on a single complex geodesic anchored at the endpoints of the timelike interval. This action consists of the complex geodesic length and the twist of a parallel-transported normal frame. We carry out the computation in Poincar\'e AdS$_3$, non-extremal and extremal rotating BTZ backgrounds, and provide an equivalent Chern-Simons derivation in which the bulk connection stays real while only the saddle of the Wilson-line probe becomes complex. In each case, the holographic result matches the corresponding TEE in dual field theory. The real part of the TEE is fixed unambiguously whereas its imaginary part depends on the boundary $i\epsilon$ prescription and the orientation of the interval. In the bulk, the dependence on orientation reflects an ambiguity in the choice of framing. Along the complex geodesic, the residual frame group becomes $SO(2,\mathbb C)\simeq\mathbb C^*$, so the endpoint conditions determine the frame twist only up to integer multiples of $2\pi i$. The difference between the two orientations, $\frac{c_L-c_R}{6}\,i\pi$, is fixed entirely by the gravitational anomaly and persists for the R\'enyi pseudo-entropies.

\end{abstract}

\end{center}

\newpage
\setcounter{footnote}{0}

\tableofcontents
\newpage

\section{Introduction}
\label{sec:intro}

Entanglement entropy has become a central quantity across quantum field theory, holography and quantum gravity. In a two-dimensional conformal field theory (CFT$_2$) the entanglement entropy of an interval is fixed by the central charge \cite{Calabrese:2004eu,Calabrese:2009qy} and holographically it is computed by the Ryu-Takayanagi formula as the length of a bulk geodesic \cite{Ryu:2006bv,Ryu:2006ef}. However when the left- and right-moving central charges differ, $c_L\neq c_R$, the theory exhibits a gravitational anomaly \cite{Alvarez-Gaume:1983ihn}: the stress tensor is not conserved, or equivalently not symmetric and the theory becomes sensitive to the choice of local frame. The entanglement entropy then depends not only on the invariant length of the interval but also on its boost. Holographically the anomaly is dual to a gravitational Chern-Simons term i.e. topologically massive gravity (TMG) \cite{Deser:1981wh,Kraus:2005zm}. The authors \cite{Castro:2014tta} showed that the geodesic is then replaced by the worldline of a massive spinning particle, so that the holographic entanglement entropy is given by the geodesic length plus the twist of a normal frame transported along it, and that the same result follows from a Wilson line in the Chern-Simons formulation of TMG \cite{Castro:2014tta,Ammon:2013hba}. The generalized Rindler method provides an alternative computation of entanglement entropy in such anomalous CFTs \cite{He:2023cju}. Various bipartite mixed state measures in anomalous CFTs have also been investigated \cite{Basu:2022nds,Wen:2022jxr}.

Most prior investigations into entanglement entropy have largely concentrated on spacelike separated subsystems. A more recent development is the timelike entanglement entropy (TEE), defined by analytically continuing the entanglement entropy of a spacelike interval to a timelike one \cite{Doi:2022iyj,Doi:2023zaf}. The TEE is complex valued in CFT$_2$ and is an example of pseudo entropy \cite{Nakata:2020luh,Mollabashi:2020yie,Mollabashi:2021xsd}. Other operational definitions of entanglement in time have been shown to coincide with it in relativistic quantum field theories \cite{Milekhin:2025ycm}. Holographically, the TEE was obtained by a combination of spacelike and timelike bulk geodesics \cite{Doi:2022iyj,Doi:2023zaf} and later by using complex-valued weak extremal surface (CWES) criteria \cite{Li:2022tsv}. For spacetimes without a known global embedding such as Lifshitz geometries, holographic TEE is computed directly in the Poincaré patch using two spacelike geodesics joined at infinity by a timelike segment \cite{Basak:2023otu}. This construction reproduces the AdS$_3$/CFT$_2$ result and has been extended to non-conformal, non-relativistic theories and black hole backgrounds using robust merging condition \cite{Afrasiar:2024lsi,Afrasiar:2024ldn,Afrasiar:2025eam}.
In \cite{Heller:2024whi}, the authors proposed that the holographic TEE is computed by a single boundary anchored extremal surface in a complexified bulk geometry, providing a geometric prescription even when no piecewise counterpart exists. Subsequently, it has been shown in \cite{Guo:2026vkl} that in locally AdS$_3$ backgrounds, the Kontsevich-Segal-Witten (KSW) criterion \cite{Kontsevich:2021dmb,Witten:2021nzp} selects the relevant complex contour uniquely.
See \cite{Wang:2018jva,Liu:2022ugc,Narayan:2022afv,Jiang:2023ffu,Narayan:2023ebn,Chu:2023zah,Jiang:2023loq,Chen:2023eic,He:2023ubi,Narayan:2023zen,Shinmyo:2023eci,Diaz:2023npx,Kanda:2023jyi,Grieninger:2023knz,Basu:2024bal,Anegawa:2024kdj,Wen:2024muv,Goswami:2024vfl,Jena:2024tly,Xu:2024yvf,Roychowdhury:2025ukl,Guo:2025pru,Roychowdhury:2025aye,Giataganas:2025div,Diaz:2025aqe,Roychowdhury:2025ebs,Heller:2025kvp,Nanda:2025tid,Zhou:2025sqq,Fujiki:2025rtx,Harper:2025lav,Giataganas:2025ize,Guo:2025dtq,Liu:2025ctb,Li:2026fcr,Prihadi:2026nua,Narayan:2026wzp,Diaz:2026qrf,Li:2026bof,Pal:2026ysc,Bernamonti:2026pxo,Bhattacharyya:2026svh,Feng:2026grc,Ladghami:2026hby,Saha:2026vin,Anastasiou:2026bbf,Kanda:2026jyk,Ali-Akbari:2026hys,Ali-Akbari:2026oez,Ori:2026lyk} for recent developments related to TEE.

More recently, \cite{Chu:2025sjv} computed the TEE in a CFT$_2$ with a gravitational anomaly both in field theory and holographically. They found that the imaginary part depends on a single central charge, so that it can serve as a probe of the anomaly. The holographic computation there uses the piecewise construction, with spacelike and timelike geodesics joined at null infinity and a prescription for the normal frames at the junctions.
These results motivate us to investigate whether anomalous TEE can be obtained using the complexified geometry prescription and the Chern–Simons formulation. In this context, we first compute the holographic TEE in a CFT$_2$ with gravitational anomaly within the complexified geometry prescription. The holographic TEE is the on-shell action of the spinning particle on a single complex geodesic anchored on the timelike interval, with the twist obtained from its closed-form expression in terms of a parallel-transported frame. No junctions and no prescription for normal frames in the bulk are needed apart from a single discrete datum discussed below, and boost and rotation of the frame become two real forms of one complexified group. 
We perform the holographic computation for TEE of a pure and boosted timelike interval in Poincar\'e AdS$_3$, and for pure timelike intervals in non-extremal and extremal rotating BTZ backgrounds following the method of \cite{Castro:2014tta}. In all cases the result reproduces the dual field theory TEE. We also discuss in what sense vectors and parallel transport are well defined in the complexified geometry and relate the contour to the KSW-admissible one.

A central point of the paper is the branch structure of the imaginary part. The holographic functional is a sum of two separately computed terms, the length and the twist, each of which produces its own $\log(-1)$. The branch of the length term is fixed by the boundary $i\epsilon$ prescription, equivalently by the orientation of the KSW contour. The branch of the twist term is fixed by a different datum, the orientation of the interval. For a timelike interval exactly one of the two light-cone separations $z_{12}$, $\bar z_{12}$ is negative, and which one depends on whether the twist operator $\Phi_n$ sits at the later or the earlier endpoint. Continuing from a spacelike interval with the standard $i\epsilon$ then gives $\frac{c_L}{6}i\pi$ in the first case and $\frac{c_R}{6}i\pi$ in the second. The result of \cite{Chu:2025sjv}, obtained by taking the principal value of each logarithm, coincides with the second. In a non-anomalous theory the two orientations agree while in the anomalous theory they differ by $\frac{c_L-c_R}{6}i\pi$, the exchange phase of the spinning twist fields. In the bulk this is a framing ambiguity of the spinning particle i.e. on a complex geodesic the boost parameter of the normal frame becomes complex and is fixed by the endpoint data only modulo $2\pi i$.

We then derive the same results in the Chern-Simons formulation following \cite{Castro:2014tta}. The anomaly appears as the asymmetry of the two Chern-Simons levels and the entropy is fixed by two exact traces of the gauge functions at the endpoints, one per chirality and the first order equations of the auxiliary probe remove the multiple saddle ambiguity of the metric formulation. On evaluating these traces at the timelike endpoints, one of the two traces falls below $-2$, the left or the right one depending on the orientation, so the probe saddle becomes complex and coincides with the complex geodesic, while the bulk connection remains real. Complexifying TMG is therefore sufficient but not necessary. We further show that evaluating the entropy formula at timelike endpoints is an analytic continuation in which the continuation data reduce to the two branch choices, and that both branches can be fixed by the boundary prescription and the orientation alone, without reference to a complex geometry.

The paper is organized as follows. Section \ref{sec:review} reviews the field-theory TEE computations in anomalous CFT$_2$ including its dependence on the orientation of the interval, together with the complexified geometry prescription and TMG. Section \ref{sec:holo} computes the holographic TEE with the complexified prescription for pure and boosted timelike intervals in Poincar\'e AdS$_3$ and in rotating BTZ, relates the contour to the KSW-admissible one, and discusses vectors and framing in the complexified geometry. Section \ref{sec:cs} gives the Chern-Simons/Wilson-line derivation, including the determination of both branches from the boundary prescription and the orientation of the Wilson line. Section \ref{sec:summary} summarizes and concludes.

\section{Review of earlier works}
\label{sec:review}

\subsection{TEE in CFT$_2$ with gravitational anomalies}
\label{sec:ft}

We begin with a brief review of two-dimensional CFTs with gravitational anomalies \cite{Alvarez-Gaume:1983ihn,Alvarez-Gaume:1984zlq}.
A two dimensional CFT with a gravitational anomaly has two copies of the Virasoro algebra with unequal central charges $c_L\neq c_R$. The energy-momentum conservation then fails at the quantum level: in the form where the stress tensor is symmetric, but not conserved which is captured by the anomalous divergence of the stress tensor \cite{Kraus:2005zm}
\begin{equation}
\nabla_\mu T^{\mu\nu}=\frac{c_L-c_R}{96\pi}\,g^{\nu\alpha}\epsilon^{\mu\beta}\partial_\mu\partial_\rho\Gamma^\rho{}_{\alpha\beta}.
\end{equation}
Equivalently, by adding a local counterterm, the anomaly can be moved into a Lorentz anomaly, with a conserved but non-symmetric stress tensor. In either way the theory is sensitive to the choice of local frame and quantities such as the entanglement entropy depend not only on the invariant length of an interval but also on its boost. We will use light cone coordinates $z=x-t$ and $\bar z=x+t$, so that left-movers depend on $z$.

\subsubsection*{Zero temperature}

\begin{figure}[H]
	\centering
	\includegraphics[scale=1]{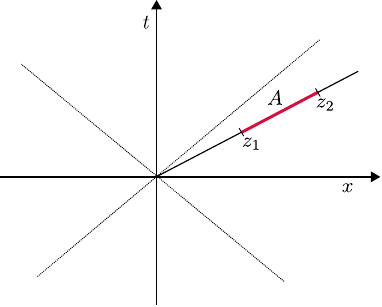}
	\caption{Schematic of a spacelike boosted interval in anomalous CFT$_2$.}\label{fig:interval}
\end{figure}

Consider a single interval $A$ with endpoints $(t_1,x_1)$ and $(t_2,x_2)$ in a CFT$_2$ with gravitational anomaly, then the entanglement entropy for $A$ can be obtained through a two-point function of twist operators $\Phi_{\pm n}$ using the replica technique \cite{Calabrese:2004eu,Calabrese:2009qy}. The configuration is shown in \cref{fig:interval}.
In the anomalous theory the twist operators carry the chiral weights as
\begin{equation}
h_n=\frac{c_L}{24}\Big(n-\frac1n\Big),\qquad \bar h_n=\frac{c_R}{24}\Big(n-\frac1n\Big),
\end{equation}
so that they have scaling dimension $\Delta_n=h_n+\bar h_n$ and a non-zero spin $s_n=h_n-\bar h_n\propto c_L-c_R$. Their two-point function is given by
\begin{equation}
\langle\Phi_n(z_1,\bar z_1)\,\Phi_{-n}(z_2,\bar z_2)\rangle=c_n\,z_{12}^{-2h_n}\,\bar z_{12}^{-2\bar h_n},
\qquad z_{12}=x_{12}-t_{12},\ \ \bar z_{12}=x_{12}+t_{12},
\end{equation}
and the replica limit gives the entanglement entropy for a spacelike interval as
\begin{equation}
S_A=\frac{c_L}{6}\log\frac{z_{12}}{\epsilon}+\frac{c_R}{6}\log\frac{\bar z_{12}}{\epsilon},
\label{eq:ft-chiral}
\end{equation}
where $\epsilon$ is a UV cutoff. Upon introducing the invariant length and the boost rapidity of the interval by
\begin{equation}
R=\sqrt{x_{12}^2-t_{12}^2},\qquad \kappa=\tanh^{-1}\frac{t_{12}}{x_{12}}=\frac12\log\frac{\bar z_{12}}{z_{12}},\label{eq:ft-Rkappa}
\end{equation}
so that $z_{12}=R\,\mathrm{e}^{-\kappa}$ and $\bar z_{12}=R\,\mathrm{e}^{\kappa}$, \eqref{eq:ft-chiral} takes the length-boost form as \cite{Castro:2014tta}
\begin{equation}
S_A=\frac{c_L+c_R}{6}\log\frac{R}{\epsilon}-\frac{c_L-c_R}{6}\,\kappa .
\label{eq:ft-lengthboost}
\end{equation}
The first term is the familiar entanglement entropy of an interval of length $R$ weighted by the average central charge.
The second term, proportional to $c_L-c_R$ captures the gravitational anomaly and reflects the non-zero spin of the twist operators.
The two terms correspond to the two separate terms of the holographic functional of \cref{sec:tmg}, the geodesic length and the frame twist. 

To continue \eqref{eq:ft-chiral} to a timelike interval we keep the labelling of the twist correlator, with $\Phi_n$ at $(t_1,x_1)$ and $\Phi_{-n}$ at $(t_2,x_2)$, and deform the interval from spacelike to timelike separation. The time of the later endpoint relative to the earlier one is continued as $t\to t(1-i\delta)$ with $\delta\to0^+$, the prescription that reproduces the TEE of \cite{Doi:2022iyj,Doi:2023zaf} for $c_L=c_R$. For $t_{12}^2>x_{12}^2$ exactly one of $z_{12}$ and $\bar z_{12}$ is negative. If $\Phi_n$ sits at the later endpoint, then $z_{12}<0<\bar z_{12}$ and $z_{12}$ approaches the negative real axis from above, so that $\log z_{12}$ acquires $+i\pi$ while $\log\bar z_{12}$ stays real. If $\Phi_n$ sits at the earlier endpoint, the roles of $z_{12}$ and $\bar z_{12}$ are exchanged. We label the two cases by $\sigma=+1$ and $\sigma=-1$, respectively, and write $c_+\equiv c_L$, $c_-\equiv c_R$. Denoting by $(t,x)$ the position of the later endpoint relative to the earlier one, with $t>|x|$, \eqref{eq:ft-chiral} becomes
\begin{equation}
S^T_{A,\sigma}=\frac{c_L}{6}\log\frac{t-x}{\epsilon}+\frac{c_R}{6}\log\frac{t+x}{\epsilon}+\frac{c_\sigma}{6}\,i\pi .
\label{eq:ft-TEE}
\end{equation}
The real part does not depend on $\sigma$. In the length-boost form \eqref{eq:ft-lengthboost} the length term acquires $\frac{c_L+c_R}{12}i\pi$ for either orientation, while the boost becomes
\begin{equation}
\kappa=\frac12\log\frac{t+x}{t-x}-\sigma\,\frac{i\pi}{2},
\label{eq:ft-kappa}
\end{equation}
so that the anomalous term contributes $\sigma\,\frac{c_L-c_R}{12}i\pi$ to the imaginary part. For a pure timelike interval $x=0$ and $t=T$, we have
\begin{equation}
S^T_{A,\sigma}=\frac{c_L+c_R}{6}\log\frac{T}{\epsilon}+\frac{c_\sigma}{6}\,i\pi .
\label{eq:ft-TEE-boxed}
\end{equation}
The result for anomalous TEE in \cite{Chu:2025sjv} was obtained differently. There one continues the length-boost form \eqref{eq:ft-lengthboost} to $t_{12}^2>x_{12}^2$, sets $x_{12}\to0^+$ with $t_{12}=T>0$, so that $\tanh^{-1}(t_{12}/x_{12})\to\frac12\log(-1)$, and takes the principal value of each logarithm. This gives $\frac{c_R}{6}i\pi$, and the same value is obtained if the endpoints are relabelled. It coincides with \eqref{eq:ft-TEE-boxed} for $\sigma=-1$. In the labelling used there, however, $\Phi_n$ sits at the later endpoint, and the limits $x\to 0^\pm$ of the boosted result \eqref{eq:ft-TEE} give $\frac{c_L}{6}i\pi$. For that labelling the principal value is therefore not the continuation in $x$.

In either orientation the real part depends on both central charges while the imaginary part depends on a single one. For $\sigma=+1$ the right-moving sector contributes only to the real part, and for $\sigma=-1$ the left-moving one does. For $c_L=c_R=c$ the two orientations agree and \eqref{eq:ft-TEE-boxed} reduces to the corresponding TEE of a non-anomalous CFT$_2$ \cite{Doi:2022iyj,Doi:2023zaf}.

The dependence on $\sigma$ reflects the spin of the twist operators. At finite $n$ the same continuation gives
\begin{equation}
S^T_{n,+}-S^T_{n,-}=\frac{2\pi i\,s_n}{n-1}=\frac{i\pi\,(c_L-c_R)(n+1)}{12n},
\label{eq:ft-orientation-gap}
\end{equation}
which vanishes for $c_L=c_R$ and reduces to $\frac{c_L-c_R}{6}i\pi$ at $n=1$. The difference between the two orientations is thus fixed by the anomaly alone. For a purely chiral theory with $c_R=0$, the algebra of a timelike segment at fixed $x$ coincides with that of a null segment, whose entanglement entropy is real. The orientation $\sigma=-1$ reproduces this, while $\sigma=+1$ gives $\frac{c_L}{6}i\pi$. For $c_L=0$ the roles are reversed, so neither orientation is preferred in general.

\subsubsection*{Finite temperature and angular potential}

In the presence of gravitational anomaly, the thermal state is a grand-canonical ensemble with a conserved spin angular momentum $J$ whose conjugate is the angular potential $\Omega$,
\begin{equation}
Z=\mathrm{Tr}\,\mathrm{e}^{-\beta H-i\beta\Omega_E J},\qquad
H=L_0+\bar L_0-\tfrac{c_L+c_R}{24},\quad J=L_0-\bar L_0-\tfrac{c_L-c_R}{24},
\end{equation}
where $\Omega_E=-i\Omega$ is the Euclidean angular potential, so that the left- and right-movers are thermal at independent temperatures $\beta_{L,R}^{-1}$ as
\begin{equation}
\beta_L=\beta(1+i\Omega_E)\;,\quad \beta_R=\beta(1-i\Omega_E)\ .
\end{equation}
The theory lives on a twisted cylinder, mapped to the plane by $z=\mathrm{e}^{2\pi w/\beta_L}$, $\bar z=\mathrm{e}^{2\pi\bar w/\beta_R}$ and the twist correlator gives the entanglement entropy of a spacelike interval. After continuing to Lorentzian signature and passing to the length-boost form as
\begin{equation}
S_A=\frac{c_L+c_R}{12}\log\!\Big[\frac{\beta_L\beta_R}{\pi^2\epsilon^2}\sinh\frac{\pi(x_{12}-t_{12})}{\beta_L}\sinh\frac{\pi(x_{12}+t_{12})}{\beta_R}\Big]
-\frac{(c_L-c_R)}{12}\log\frac{\beta_R\sinh\frac{\pi(x_{12}+t_{12})}{\beta_R}}{\beta_L\sinh\frac{\pi(x_{12}-t_{12})}{\beta_L}} ,
\end{equation}
where the second term is again the anomalous contribution. For a timelike interval, exactly one of the two $\sinh$ factors has a negative argument, the first for $\sigma=+1$ and the second for $\sigma=-1$, and with the prescription above its logarithm acquires $+i\pi$. Setting $x=0$ and $t=T$ gives
\begin{equation}
S^T_{A,\sigma}=\frac{c_L+c_R}{12}\log\!\Big[\frac{\beta_L\beta_R}{\pi^2\epsilon^2}\sinh\frac{\pi T}{\beta_L}\sinh\frac{\pi T}{\beta_R}\Big]
-\frac{(c_L-c_R)}{12}\log\frac{\beta_R\sinh\frac{\pi T}{\beta_R}}{\beta_L\sinh\frac{\pi T}{\beta_L}}
+\frac{c_\sigma}{6}\,i\pi .
\label{eq:ft-TEE-btz}
\end{equation}
Regrouping the real part leads to the following
\begin{equation}
S^T_{A,\sigma}=\frac{c_L}{6}\log\Big[\frac{\beta_L}{\pi\epsilon}\sinh\frac{\pi T}{\beta_L}\Big]+\frac{c_R}{6}\log\Big[\frac{\beta_R}{\pi\epsilon}\sinh\frac{\pi T}{\beta_R}\Big]+\frac{c_\sigma}{6}\,i\pi .
\end{equation} The imaginary part is the same as at zero temperature, while the anomaly now also contributes to the real part. In the limit $\beta_{L,R}\gg T$, \eqref{eq:ft-TEE-btz} reduces to \eqref{eq:ft-TEE-boxed}.

\subsubsection*{Zero temperature and finite angular potential}

At zero temperature with finite angular momentum, the left-movers are in their ground state $\beta_L\to\infty$, while the right-movers remain at an effective (Frolov-Thorne) temperature $\beta_R^{-1}=T_{FT}=r_0/\pi\ell$ \cite{PhysRevD.39.2125}, the parameter $r_0$ being the horizon radius of the dual extremal BTZ black hole. The map from the plane is $z=w$, $\bar z=\frac{\ell}{2r_0}\mathrm{e}^{2r_0\bar w/\ell}$, and the entanglement entropy of a spacelike interval is a zero-temperature result for the left-movers and a thermal one at $T_{FT}$ for the right-movers. In length-boost form, the entanglement entropy for this case can be written as
\begin{equation}
\begin{aligned}
S_A={}&\frac{c_L+c_R}{12}\log\!\Big[\frac{\ell\,(x_{12}-t_{12})}{r_0\epsilon^2}\sinh\frac{r_0(x_{12}+t_{12})}{\ell}\Big]\\
&-\frac{(c_L-c_R)}{12}\log\Big[\frac{\ell}{r_0(x_{12}-t_{12})}\sinh\frac{r_0(x_{12}+t_{12})}{\ell}\Big].
\end{aligned}
\end{equation}
Now continuing to a pure timelike interval as before gives
\begin{equation}
\begin{aligned}
S^T_{A,\sigma}&=\frac{c_L+c_R}{12}\log\!\Big[\frac{\ell T}{r_0\epsilon^2}\sinh\frac{r_0T}{\ell}\Big]
-\frac{(c_L-c_R)}{12}\log\Big[\frac{\ell}{r_0T}\sinh\frac{r_0T}{\ell}\Big]+\frac{c_\sigma}{6}\,i\pi\\
&=\frac{c_L}{6}\log\frac{T}{\epsilon}+\frac{c_R}{6}\log\Big[\frac{\ell}{r_0\epsilon}\sinh\frac{r_0T}{\ell}\Big]+\frac{c_\sigma}{6}\,i\pi .
\end{aligned}
\label{eq:ft-TEE-ext}
\end{equation}
The left-moving sector contributes its zero-temperature part and the right-moving sector the thermal factor at $T_{FT}$. For $\sigma=-1$ the right-moving sector also carries the entire imaginary part and for $\sigma=+1$ the imaginary part $\frac{c_L}{6}i\pi$ is carried by the left-moving sector which is otherwise in its ground state. This is the $\beta_L\to\infty$ limit of \eqref{eq:ft-TEE-btz}.

\subsection{Holographic TEE using complexified prescription}
\label{sec:heller}

The holographic entanglement entropy of a spacelike subsystem is the area (length) of a boundary anchored extremal surface (geodesic) in the real Lorentzian bulk. For a timelike interval no such real extremal surface exists. The authors \cite{Heller:2024whi} proposed that the holographic timelike entanglement entropy is given by a boundary anchored extremal surface $\gamma_A$, now living in a complexified bulk geometry as
\begin{equation}
S_{A}^T=\frac{\mathrm{Area}(\gamma_A)}{4G},
\end{equation}
where complexified geometry means the bulk coordinates are promoted to complex variables while the asymptotic boundary is kept real. The surface is anchored on the real timelike interval and in general extends into the complex bulk. Its area is complex and reproduces the TEE of dual field theory for case of CFT$_2$. Upon analytic continuation to a spacelike interval the prescription reduces to ordinary RT formula.

For a pure timelike interval $A=[-T/2,T/2]$ at $x=0$ in Poincar\'e AdS$_3$, the holographic configuration is a single complex geodesic with a complex turning point (tip) at radial coordinate $u_t=iT/2$. This is the analytic continuation of the real spacelike RT family into complex radius. A useful feature of this construction is that special contours in the complex affine parameter plane pass through distinct real sections of the complexified geometry where two segments correspond to the spacelike branches and an intermediate imaginary segment generates the imaginary contribution to the TEE.
This construction as originally stated is for Einstein gravity where the extremal surface contributes only through its length (or area). To obtain the anomalous holographic TEE, we must combine it with the TMG spinning-particle functional reviewed below. The relevant bulk probe is therefore a massive spinning particle whose worldline is the complex geodesic and whose on-shell action is the sum of the geodesic length and a Chern-Simons frame twist.

\subsection{TMG-AdS$_3$}
\label{sec:tmg}

For conformal field theories with a gravitational anomaly, the bulk dual geometry is described by topologically massive gravity
(TMG) in a bulk AdS$_3$ spacetime whose action is given by \cite{Kraus:2005zm, Tachikawa:2006sz,Skenderis:2009nt,Castro:2014tta}
\begin{equation}\label{eq:TMG-action}
S_{\rm TMG}=\frac{1}{16\pi G^{(3)}_N}\int \mathrm{d}^3x\,\sqrt{-g}\Big(R+\frac{2}{\ell^2}\Big)
+\frac{1}{32\pi G^{(3)}_N\,\mu}\int \mathrm{d}^3x\,\sqrt{-g}\,\epsilon^{\lambda\mu\nu}
\Gamma^{\rho}{}_{\lambda\sigma}\Big(\partial_\mu\Gamma^{\sigma}{}_{\rho\nu}
+\tfrac{2}{3}\Gamma^{\sigma}{}_{\mu\tau}\Gamma^{\tau}{}_{\nu\rho}\Big),
\end{equation}
where the first term represents the standard Einstein-Hilbert action and second one is gravitational Chern-Simons (CS) term. Here $\mu$ is the dimensionful coupling associated with the Chern–Simons term while $\ell$ denotes the AdS$_3$ radius. The Brown-Henneaux \cite{Brown:1986nw} type central charges for TMG-AdS$_3$ is given by \cite{Hotta:2008yq,Compere:2008cv}
\begin{equation}
c_L=\frac{3\ell}{2G^{(3)}_N}\Big(1-\frac{1}{\mu\ell}\Big),\qquad
c_R=\frac{3\ell}{2G^{(3)}_N}\Big(1+\frac{1}{\mu\ell}\Big),
\quad\mathrm{or}\quad
\frac{1}{4G^{(3)}_N\,\mu}=\frac{c_R-c_L}{12},\quad
\frac{\ell}{4G^{(3)}_N}=\frac{c_L+c_R}{12}.
\label{eq:centralcharges}
\end{equation}
We use boundary light-cone coordinates $z=x-t$, $\bar z=x+t$, so that left-movers are functions of $z$. We set $\ell=1$ except where its restoration is instructive.

As described in \cite{Castro:2014tta}, the holographic entanglement entropy in TMG is computed by the on-shell action of a massive spinning particle as
\begin{equation}
S =\frac{1}{4G^{(3)}_N}\,\underset{\gamma}{\rm ext}
\left[\,\int_\gamma d\lambda\,\sqrt{g_{\mu\nu}\dot X^\mu\dot X^\nu}
+\frac{1}{\mu}\int_\gamma d\lambda\;\tilde n\cdot\nabla n\,\right]
\;\equiv\;S_{\rm EH}+S_{\rm anom},
\label{eq:onshell}
\end{equation}
where $(n,\tilde n)$ are normal vectors to the worldline $\gamma$ and $\nabla$ is the covariant derivative along it. In locally AdS$_3$ solutions the extremal worldlines are geodesics and the Chern-Simons contribution is a boundary term. Upon introducing a parallel-transported normal frame $(q,\tilde q)$ with
\begin{equation}
v^2=1,\quad q^2=-1,\quad \tilde q^2=1,\quad
v\cdot q=v\cdot\tilde q=q\cdot\tilde q=0,\quad
\nabla v=\nabla q=\nabla\tilde q=0,
\label{eq:frameconds}
\end{equation}
the anomalous contribution to the holographic entanglement entropy can be obtained as \cite{Castro:2014tta}
\begin{equation}
S_{\rm anom}= \frac{1}{4 G_N^{(3)}\mu} \left(\eta(s_f) -
\eta(s_i)\right)=\frac{1}{4G_N^{(3)}
  \mu}\log\left(\frac{q(s_f) \cdot n_f - \tilde{q}(s_f) \cdot
  n_f}{q(s_i) \cdot n_i - \tilde{q}(s_i) \cdot n_i}\right).
\label{eq:twist}
\end{equation}
Here $n_i$ and $n_f$ are normal vectors at the end point of worldline, given by 
\begin{equation}
    n(s_i) = n_i ,\quad n(s_f) = n_f.
\end{equation}
We orient the worldline from $X_i=X(s_i)$ to $X_f=X(s_f)$. For a spacelike interval \eqref{eq:twist} then reproduces the anomalous term of \eqref{eq:ft-lengthboost} with $\eta_f-\eta_i=2\kappa$, where $\kappa$ is the boost \eqref{eq:ft-Rkappa} with $\Phi_n$ at $X_f$ and $\Phi_{-n}$ at $X_i$, see \cref{sec:boosted}.

\section{Holographic computation of anomalous TEE using the complexified prescription}
\label{sec:holo}

We now compute the anomalous TEE holographically, using the complexified-geometry prescription of \cref{sec:heller} together with the TMG spinning-particle functional of \cref{sec:tmg}. We first contrast the piecewise and complexified prescriptions, then carry out the vacuum and rotating BTZ computations, discuss the advantages of the complexified prescription, its connection to KSW admissibility and the meaning of vectors in the complexified geometry.

\subsection{The two prescriptions for holographic TEE}

\subsubsection*{Piecewise prescription} 

In the piecewise prescription \cite{Chu:2025sjv}, the holographic TEE is obtained by extremizing the sum of the on-shell actions associated with the spacelike and timelike geodesic as follows
\begin{equation}\label{TEE-total-expression}
  S_{\rm TEE}=\mathrm{ext}_C ~
  \left(S_{\rm on-shell}^{\rm spacelike}+S_{\rm on-shell}^{\rm timelike}\right).
\end{equation}
The above expression contains both the geodesic length and the anomalous contribution corresponding to the spacelike and timelike geodesic.
The Einstein-Hilbert term gives the length of spacelike and timelike geodesics while the Chern-Simons term contributes a boost (twist) along the spacelike geodesic as given in \eqref{eq:twist} and a rotation $\theta$ along the timelike geodesic which is given by \cite{Chu:2025sjv}
\begin{equation}\label{eq:twistrot}
S_{\rm anom}^{\rm timelike} = \frac{i}{4 G_N^{(3)}\mu} \left(\theta(\tau_f)
  - \theta(\tau_i)\right)=\frac{1}{4G_N^{(3)} \mu}\log \left[\frac{q(\tau_f)\cdot n_f-i \,\tilde{q}(\tau_f)\cdot n_f}{q(\tau_i)\cdot n_i-i \,\tilde{q}(\tau_i)\cdot n_i}\right],
\end{equation}
where the parallel transported vectors used above satisfy the following condition
\begin{equation}\label{Timelike-q-condition}
 \begin{aligned}
   v^2= -1,\quad q^2=\tilde{q}^2=1,\quad q \cdot
   \tilde{q} =v\cdot q =v\cdot \tilde{q} =0,
\quad\nabla v = 0\ , \quad \nabla q=0\ , \quad \nabla \tilde{q} = 0.
 \end{aligned}
\end{equation}
A prescription for the normal vectors at the endpoints of these geodesics, both on the boundary and deep in the bulk at null infinity was proposed in \cite{Chu:2025sjv} and reproduces the dual field theory result. The bulk part of this prescription is an additional input, since the junctions at null infinity are not fixed by the boundary data. In the complexified prescription below, the normal vector is specified only at the real boundary endpoints and the remaining discrete datum is the orientation of the worldline.

\subsubsection*{Complexified prescription}

In this proposal, the holographic TEE is given by a single codimension two extremal surface (geodesic) $\gamma_A$ anchored on the real timelike subregion and extending into the complexified bulk where all coordinates are promoted to complex variables. The holographic TEE in TMG-AdS$_3$ is then given by the on-shell action \eqref{eq:onshell} evaluated on a single complex geodesic $\gamma_A$ anchored on the real timelike interval $A$, with the tangent $v$, the parallel transported vectors $(q,\tilde q)$ and the normal vector $n$ all analytically continued. It can then be expressed as
\begin{equation}
S_{\rm TEE} =\frac{1}{4G^{(3)}_N}\,\underset{\gamma}{\rm ext}
\left[\,\int_\gamma d\lambda\,\sqrt{g_{\mu\nu}\dot X^\mu\dot X^\nu}
+\frac{1}{\mu}\int_\gamma d\lambda\;\tilde n\cdot\nabla n\,\right],
\end{equation}
where the first term corresponds to the length of the geodesic and second corresponds to the twist along the geodesic in complexified spacetime. The boundary conditions are imposed only at the real boundary endpoints, $n=\partial_t$ (normalized) at both ends of $A$. The twist is computed from \eqref{eq:twist} with complex $\eta$ or $(n,q,\tilde q)$. Since the normal frame is now complex, $\eta$ takes values in $\mathbb C$ and the endpoint data fix $\eta_f-\eta_i$ only modulo $2\pi i$. We return to this in \cref{sec:framing}.

So, we observe the unification of boost and rotation in the anomalous theory where 
\eqref{eq:twist} with $\eta\in\mathbb{C}$ contains both \eqref{eq:twist} with real $\eta$ (spacelike segments) and \eqref{eq:twistrot} (timelike segments): for real spacelike curves $\eta$ is a boost parameter of $SO(1,1;\mathbb{R})$, for real timelike curves $\eta=i\theta$ is a rotation angle of $SO(2;\mathbb{R})$, and on a complex geodesic $\eta$ parametrizes the complexified residual frame group
\begin{equation}
SO(1,1;\mathbb{C})\;\simeq\;SO(2;\mathbb{C}),
\end{equation}
of which the boost and rotation groups are the two real forms. In this case no junction conditions at null infinity and no large-rapidity regularization are needed: the frame is parallel-transported holomorphically through the complex bulk, and the only remaining datum is the orientation of the worldline.
We now apply this complexified prescription to compute the holographic TEE in several cases discussed in the following subsections.

\subsection{Vacuum: Poincar\'e AdS$_3$}

\begin{figure}[h]
	\centering
	\includegraphics[scale=1]{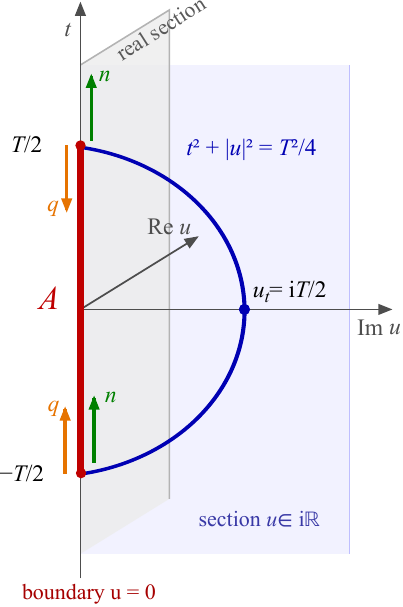}
	\caption{Schematic of a single complex geodesic for computing the anomalous holographic TEE of a pure timelike interval $A$. The radial coordinate is complexified, with $\mathrm{Im}\,u$ in the plane of the page and $\mathrm{Re}\,u$ oblique, the gray sheet is the real Lorentzian section and the $x$ direction is suppressed. Along real values of the affine parameter, the geodesic \eqref{eq:complexgeo} has $t=\tfrac{T}{2}\tanh\lambda\in\mathbb{R}$ and $u=\tfrac{iT}{2}\,{\rm sech}\,\lambda\in i\mathbb{R}$, so it is the semicircle $t^2+|u|^2=T^2/4$ with tip $u_t=iT/2$. At the endpoints, the boundary normal is $n=\partial_t$ (green), and the parallel-transported frame vector satisfies $q=+n$ at $t=-T/2$ and $q=-n$ at $t=+T/2$ (orange), giving the twist ratio $-1$.}\label{fig:complex-geodesic}
\end{figure}

Consider a pure timelike interval $A\equiv[-\frac{T}{2},\frac{T}{2}]$ of length $T$ in a dual CFT$_2$ with a gravitational anomaly at zero temperature. The corresponding holographic dual is described by the AdS$_3$ spacetime in Poincar\'e
coordinates as
\begin{equation}
ds^2=\frac{1}{u^2}(-dt^2 + dx^2 +du^2).
\label{eq:poincare}
\end{equation}
The complex geodesic at leading order in the cutoff $\epsilon$ is given by \cite{Heller:2024whi}
\begin{equation}
t(\lambda)=\frac{T}{2}\tanh\lambda,\qquad
u(\lambda)=\frac{i\,T}{2}\,{\rm sech}\,\lambda,\qquad
\lambda\in[-\lambda_*,\lambda_*],\qquad
\lambda_*=\log\frac{T}{\epsilon}+\frac{i\pi}{2}.
\label{eq:complexgeo}
\end{equation}
This complex geodesic structure is depicted in \cref{fig:complex-geodesic}.

\subsubsection*{Length contribution}

With affine normalization $\mathcal{L}=1$, the regularized length is given by $L=2\lambda_*$, so
\begin{equation}
S_{\rm EH}=\frac{L}{4G^{(3)}_N}
=\frac{c_L+c_R}{6}\log\frac{T}{\epsilon}
+\frac{c_L+c_R}{12}\,i\pi.
\label{eq:SEHvac}
\end{equation}
This equals the sum of the two real spacelike lengths and the $i\pi$ of the timelike segment in the piecewise computation \cite{Chu:2025sjv} but here both arise from one analytic curve.

\subsubsection*{Anomaly contribution}
\label{sec:twist}

The orthonormal frame of the real spacelike family analytically continues as the defining conditions \eqref{eq:frameconds} are polynomial in the components, hence hold for complex coordinates. So, the tangent vector and
parallel transported vector can be written as (in $(t,x,u)$ convention)
\begin{equation}
v^\mu=-\frac{2u}{T}\,(u,0,t),\qquad
q^\mu=-\frac{2u}{T}\,(t,0,u),\qquad
\tilde q^\mu=(0,u,0),
\label{eq:frame}
\end{equation}
and the (normalized, future-pointing) boundary value of normal vector continues to $n^\mu=(u,0,0)|_{u=\epsilon}$ satisfying $n^2=g_{tt}(n^t)^2=-1$ and reducing to $\partial_t$ at the boundary. Here the initial and final value of normal vector $n$ at the boundary is same as considered in \cite{Castro:2014tta} which is determined by the boundary CFT data.
On using \eqref{eq:frame} with $n_i=n_f=\partial_t$, the ratio in \eqref{eq:twist} is given by
\begin{equation}
\frac{\big(q\cdot n-\tilde q\cdot n\big)\big|_f}{\big(q\cdot n-\tilde q\cdot n\big)\big|_i}=-1,
\end{equation}
so that the endpoints data give $\eta_f-\eta_i=\pm i\pi$ but do not decide the sign. We fix it as on the field theory side, by continuing from a spacelike interval and the explicit continuation is given for general intervals in \cref{sec:boosted}. The result is $\eta_f-\eta_i=-\sigma\,i\pi$, with $\sigma=+1$ if $X_f$ is the later endpoint and $\sigma=-1$ otherwise. Hence we have
\begin{equation}
S_{\rm anom}=-\sigma\,\frac{i\pi}{4G^{(3)}_N\mu}=\sigma\,\frac{c_L-c_R}{12}\,i\pi .
\label{eq:anomvac}
\end{equation}

The holographic TEE is the sum of the length and anomaly on-shell actions \eqref{eq:onshell}. On using \eqref{eq:SEHvac} and \eqref{eq:anomvac},  the
holographic TEE in the presence of gravitational anomaly can be written as
\begin{equation}\label{eq:mainvac}
\begin{aligned}
S^T_{A,\sigma}&=S_{\rm EH}+S_{\rm anom}
=\frac{c_L+c_R}{6}\log\frac{T}{\epsilon}
+\frac{c_L+c_R}{12}\,i\pi
+\sigma\,\frac{c_L-c_R}{12}\,i\pi\\
&=\frac{c_L+c_R}{6}\,\log\frac{T}{\epsilon}+\frac{c_\sigma}{6}\,i\pi .
\end{aligned}
\end{equation}
This reproduces the field-theory result \eqref{eq:ft-TEE-boxed} for both orientations. The length contributes $\frac{c_L+c_R}{12}i\pi$, symmetric in the central charges and independent of $\sigma$, while the twist contributes $\sigma\frac{c_L-c_R}{12}i\pi$. The worldline \eqref{eq:complexgeo}, oriented from $t=-T/2$ to $t=T/2$ as in \eqref{eq:frame}, has $\sigma=+1$ and gives $\frac{c_L}{6}i\pi$. The opposite orientation gives $\frac{c_R}{6}i\pi$, which is the result obtained in \cite{Chu:2025sjv} with the piecewise construction. In the non-anomalous limit $c_L=c_R=c$ the twist term vanishes for either orientation and \eqref{eq:mainvac} reduces to the standard holographic TEE \cite{Doi:2022iyj,Doi:2023zaf}.

\subsection{Boosted timelike intervals}
\label{sec:boosted}

The above computation extends directly to a general timelike interval and this is also where the orientation enters explicitly. Let the endpoints $X_i$ and $X_f$ lie on the boundary of \eqref{eq:poincare} with $t_{fi}^2>x_{fi}^2$ and $R_P=\sqrt{t_{fi}^2-x_{fi}^2}$. The complex geodesic is then the boost of \eqref{eq:complexgeo} as
\begin{equation}
t-t_c=\frac{t_{fi}}{R_P}\,\hat\lambda,\qquad
x-x_c=\frac{x_{fi}}{R_P}\,\hat\lambda,\qquad
\hat\lambda=\frac{R_P}{2}\tanh s,\qquad
u=\frac{iR_P}{2}\,{\rm sech}\,s,
\label{eq:boostedgeo}
\end{equation}
where $(t_c,x_c)$ is the midpoint and the tip sits at $u_t=i R_P/2$. Its regularized length is $L=2\log(R_P/\epsilon)+i\pi$. The tangent and the parallel-transported frame are the boosts of \eqref{eq:frame}, given by
\begin{equation}
v^\mu=-\frac{2u}{R_P}\Big(\frac{t_{fi}}{R_P}u,\ \frac{x_{fi}}{R_P}u,\ \hat\lambda\Big),\qquad
q^\mu=-\frac{2u}{R_P}\Big(\frac{t_{fi}}{R_P}\hat\lambda,\ \frac{x_{fi}}{R_P}\hat\lambda,\ u\Big),\qquad
\tilde q^\mu=u\Big(\frac{x_{fi}}{R_P},\ \frac{t_{fi}}{R_P},\ 0\Big),
\label{eq:boostedframe}
\end{equation}
which satisfy \eqref{eq:frameconds} since the conditions are Lorentz invariant. With $n^\mu=(u,0,0)$ at the two endpoints
\begin{equation}
\big(q\cdot n-\tilde q\cdot n\big)\big|_i=\frac{x_{fi}-t_{fi}}{R_P},\qquad
\big(q\cdot n-\tilde q\cdot n\big)\big|_f=\frac{x_{fi}+t_{fi}}{R_P},
\end{equation}
so that $\eta_f-\eta_i=\log(\bar z_{fi}/z_{fi})$. For a spacelike interval this is $2\kappa$, with $\kappa$ the boost \eqref{eq:ft-Rkappa} of the interval with $\Phi_n$ at $X_f$. Continuing with the prescription of \cref{sec:ft}, $z_{fi}$ crosses zero from above when $X_f$ is the later endpoint, and $\bar z_{fi}$ does so when $X_f$ is the earlier one. On writing $(t,x)$ for the position of the later endpoint relative to the earlier one, we find
\begin{equation}
\eta_f-\eta_i=\log\frac{t+x}{t-x}-\sigma\,i\pi,\qquad \sigma={\rm sign}(t_f-t_i),
\label{eq:twist-boosted}
\end{equation}
and holographic TEE can then be obtained as
\begin{equation}
S^T_{A,\sigma}=\frac{c_L+c_R}{6}\log\frac{\sqrt{t^2-x^2}}{\epsilon}+\frac{c_R-c_L}{12}\log\frac{t+x}{t-x}+\frac{c_\sigma}{6}\,i\pi ,
\end{equation}
which matches with dual field theory result \eqref{eq:ft-TEE}. The result is continuous in $x$ and the limits $x\to0^\pm$ reproduce \eqref{eq:mainvac} with the same $\sigma$. The anomalous term now also contributes to the real part, which gives a further check of the twist beyond its imaginary part.

\subsection{Finite temperature and angular potential: rotating BTZ}
\label{sec:btz}

A CFT$_2$ with a gravitational anomaly at finite temperature and finite angular potential is a grand-canonical ensemble whose conserved charges are the energy and the (spin) angular momentum. Its holographic dual is described by the rotating BTZ black brane with the metric
\begin{equation}
 ds^2 = -\frac{(r^2 - r_+^2)(r^2 - r_-^2)}{r^2\ell^2} d\tau^2
  + \frac{r^2 \ell^2}{(r^2 - r_+^2)(r^2 - r_-^2)} dr^2
  + r^2 (d\phi + \frac{r_+ r_-}{r^2\ell }d\tau)^2 ,
\label{eq:btzmetric}
\end{equation}
with outer and inner horizons at $r=r_+$ and $r=r_-$ ($r_+\ge r_-\ge 0$). The Hawking temperature and the left/right moving temperatures are given by
\begin{equation}
T_H=\frac{r_+^2-r_-^2}{2\pi r_+\ell^2},\qquad
T_{L,R}=\frac{1}{\beta_{L,R}}=\frac{r_+\mp r_-}{2\pi\ell},
\label{eq:btztemps}
\end{equation}
The dual CFT lives on a twisted cylinder and the left- and right-moving sectors are thermal at the two independent temperatures $T_L$ and $T_R$ which is characteristic of the conserved spin.

Since the rotating BTZ black hole is locally equivalent to pure AdS$_3$, all local data like geodesic lengths, parallel transport and the frame twist can be computed by mapping \eqref{eq:btzmetric} to Poincar\'e AdS$_3$,
\begin{equation}
ds^2=\frac{\ell^2}{u^2} (-dt^2+dx^2+du^2),
\end{equation}
via the coordinate map
\begin{equation}
x \pm t = \sqrt{\frac{r^2 - r_+^2}{r^2 - r_-^2}} e^{2\pi (\phi \pm \tau)/\beta_{R,L}},
  \,\, \qquad u = \sqrt{\frac{r_+^2 - r_-^2}{r^2 - r_-^2}}
  e^{(\phi r_+ + \tau r_-)/\ell}.
\label{eq:btzmap}
\end{equation}

In the complexified prescription, the holographic TEE is obtained by a single complex geodesic, so no junction of spacelike and timelike piece and no prescription for normal vectors at the junctions as discussed in \cite{Chu:2025sjv}, is needed. The map \eqref{eq:btzmap} sends the complex geodesic anchored on the BTZ interval to the (boosted and translated) complex geodesic of Poincar\'e AdS$_3$ anchored on the images of the two endpoints. Its length and its parallel-transported frame $(q,\tilde q)$ are therefore those of the vacuum computation given by \eqref{eq:boostedgeo} and \eqref{eq:boostedframe}.

Let us now consider a pure timelike interval $A$ of length $T$ in a dual CFT$_2$ at finite temperature and angular potential with endpoints,
\begin{equation}
(\tau_1,\phi_1)=\Big(-\tfrac{T}{2},0\Big),\qquad
(\tau_2,\phi_2)=\Big(\tfrac{T}{2},0\Big).
\end{equation}
Their images on the Poincar\'e boundary follow from \eqref{eq:btzmap} at the cutoff surface and form a boosted timelike interval with
\begin{equation}
\bar z_{21}=2\sinh\frac{\pi T}{\beta_R},\qquad z_{21}=-2\sinh\frac{\pi T}{\beta_L},\qquad
R_P^2=-z_{21}\bar z_{21}=4\sinh\frac{\pi T}{\beta_L}\sinh\frac{\pi T}{\beta_R}
\end{equation}
to leading order. The map preserves the boundary time orientation, so $X_2$ is also the later endpoint of the image interval. The complex geodesic anchored on the image endpoints is \eqref{eq:boostedgeo} with $(t_{fi},x_{fi})=(t_{21},x_{21})$.

\subsubsection*{Length contribution}

Upon regulating at $u=\epsilon_1$ and $u=\epsilon_2$ near the two endpoints leads to $s_{1,2}=\mp\big[\log(R_P/\epsilon_{1,2})+\tfrac{i\pi}{2}\big]$, hence the length of the geodesic in Poincar\'e coordinates is given by
\begin{equation}\label{eq:L-BTZ-rotating}
L=s_2-s_1=\log\frac{R_P^2}{\epsilon_1 \epsilon_2}+i\pi ,
\end{equation}
where the relation between the cut-off $\epsilon_{1,2}$ and $r_\infty$ can be obtained using \eqref{eq:btzmap} as
\begin{equation}\label{cut-off-relation}
\varepsilon_{i}= \sqrt{\frac{r_{+}^2-r_{-}^2}{r_{\infty}^2}}e^{\tau_i
  r_{-}/\ell}.
\end{equation}
Now using \eqref{eq:btzmap} and \eqref{cut-off-relation} in \eq{eq:L-BTZ-rotating}, the length
contribution to holographic TEE in terms of the original rotating BTZ
coordinates can be obtained as
\begin{equation}
S_{\rm EH}=\frac{\ell}{4G^{(3)}_N}\log\!\left[\frac{\beta_L\beta_R}{\pi^2\epsilon^2}\,\sinh\frac{\pi T}{\beta_L}\,\sinh\frac{\pi T}{\beta_R}\right]
+\frac{\ell}{4G^{(3)}_N}\,i\pi \ ,
\label{eq:btzSEH}
\end{equation}
where $\epsilon$ is the boundary cutoff and is related to the bulk radial cutoff $r_\infty$ by $\epsilon=\ell/r_\infty$.

\subsubsection*{Anomaly contribution}
 
Now for the anomalous contribution, the tangent and the parallel-transported frame are given by \eqref{eq:boostedframe} with $(t_{fi},x_{fi})=(t_{21},x_{21})$. The boundary conditions on the normal vector is given by the vector along the BTZ time direction \cite{Castro:2014tta}
\begin{equation}\label{eq:BC-BTZ}
    n_1 = n_2 = \left(\partial_{\tau}\right)_{\rm CFT}.
\end{equation}
The anomalous contribution to holographic TEE may now be obtained using \eq{eq:boostedframe}, \eqref{eq:btzmap} and \eq{eq:BC-BTZ} in \eq{eq:twist} as follows
\begin{equation}
S_{\rm anom}=\frac{1}{4G^{(3)}_N\mu}\log\!\left[\frac{\beta_R\,\sinh\frac{\pi T}{\beta_R}}{\beta_L\,\sinh\frac{\pi T}{\beta_L}}\right]
-\sigma\frac{1}{4G^{(3)}_N\mu}\,i\pi \ ,
\label{eq:btzSanom}
\end{equation}
where the imaginary part follows from \eqref{eq:twist-boosted} with $\sigma=+1$ for the worldline running from $X_1$ to $X_2$. At each endpoint the boundary normal $(\partial_\tau)_{\rm CFT}$ is a future-directed timelike vector that differs from $\partial_t$ of the Poincar\'e patch by a real boost. It therefore changes only the real part of $\eta_f-\eta_i$, producing the first term of \eqref{eq:btzSanom}.
Now using \eq{eq:btzSEH}, \eq{eq:btzSanom} and \eqref{eq:centralcharges}, the holographic TEE at finite temperature with a conserved angular momentum can be written as
\begin{equation}
S^T_{A,\sigma}
=\frac{c_L+c_R}{12}\log\!\left[\frac{\beta_L\beta_R}{\pi^2\epsilon^2}\,\sinh\frac{\pi T}{\beta_L}\sinh\frac{\pi T}{\beta_R}\right]
+\frac{c_R-c_L}{12}\log\!\left[\frac{\beta_R\,\sinh\frac{\pi T}{\beta_R}}{\beta_L\,\sinh\frac{\pi T}{\beta_L}}\right]
+\frac{c_\sigma}{6}\,i\pi.
\label{eq:btzresult}
\end{equation}
This agrees with the field-theory result \eqref{eq:ft-TEE-btz} for both orientations, and for $\sigma=-1$ with the piecewise holographic result of \cite{Chu:2025sjv}.

\subsection{Zero temperature and finite angular potential: extremal BTZ}
\label{sec:extremal}

The extremal rotating BTZ black hole is the limit $r_+=r_-\equiv r_0$, in which the outer and inner horizons coincide and the Hawking temperature vanishes, $T_H\to0$. The metric \eqref{eq:btzmetric} becomes
\begin{equation}
ds^{2}=-\frac{(r^2-r_{0}^2)^2}{r^2\ell^2}d\tau^2 + \frac{r^2 \ell^2
 }{(r^2-r_{0}^2)^2}dr^2+ r^2(d\phi+\frac{r_{0}^2}{r^2\ell }d\tau)^2.
\label{eq:extmetric}
\end{equation}
From \eqref{eq:btztemps}, the extremal limit is $\beta_L\to\infty$ i.e the left-movers are in their ground state while the right-movers remain at a finite effective temperature known as the Frolov-Thorne temperature \cite{PhysRevD.39.2125}
\begin{equation}
T_L=\frac{1}{\beta_L}\to0,\qquad
T_{FT}=T_R=\frac{1}{\beta_R}=\frac{r_0}{\pi\ell}.
\label{eq:FT}
\end{equation}
The Frolov-Thorne temperature represents the degeneracy of the ground state of an extremal BTZ black hole. The extremal geometry is again locally AdS$_3$ but the appropriate coordinate map degenerates from \eqref{eq:btzmap} because $r_+-r_-\to0$. The correct extremal map (the $r_+\to r_-$ limit of the non-extremal one) is given by \cite{Keski-Vakkuri:1998gmz}
\begin{equation}
x+t=\frac{\ell}{2r_0}\,\mathrm{e}^{\frac{2r_0}{\ell}(\phi+\tau)},
\qquad
x-t=\phi-\tau-\frac{\ell\,r_0}{r^2-r_0^2},
\qquad
u=\frac{\ell}{\sqrt{r^2-r_0^2}}\,\mathrm{e}^{\frac{r_0}{\ell}(\phi+\tau)} .
\label{eq:extmap}
\end{equation}
Under the map \eqref{eq:extmap}, the right-moving combination $\phi+\tau$ enters through an exponential that is a thermal map at temperature $T_{FT}$, while the left-moving combination $\phi-\tau$ enters linearly i.e. through a zero-temperature, vacuum like map. This suggests that the left sector is in its ground state and the right sector is thermal.

\subsubsection*{Length and anomaly contributions}

The computations for this case are similar to \cref{sec:btz} where we utilize the map \eqref{eq:extmap} instead of \eqref{eq:btzmap}. The image interval has $z_{21}=-T$ and $\bar z_{21}=\frac{\ell}{r_0}\sinh\frac{r_0T}{\ell}$, so $R_P^2=\frac{\ell T}{r_0}\sinh\frac{r_0T}{\ell}$. Hence the length of the single complex geodesic $L=\log(R_P^2/\epsilon_1 \epsilon_2)+i\pi$ leads to the following
\begin{equation}
S_{\rm EH}=\frac{\ell}{4G^{(3)}_N}\log\!\left[\frac{\ell T}{r_0\epsilon^2}\,\sinh\frac{r_0 T}{\ell}\right]+\frac{\ell}{4G^{(3)}_N}\,i\pi .
\end{equation}
Similarly, the anomalous contribution can be obtained by using \eq{eq:boostedframe}, \eqref{eq:extmap} and \eq{eq:BC-BTZ} in \eq{eq:twist} as follows
\begin{equation}
S_{\rm anom}=\frac{1}{4G^{(3)}_N\mu}\log\!\left[\frac{\ell}{r_0 T}\sinh\frac{r_0 T}{\ell}\right]-\sigma\frac{1}{4G^{(3)}_N\mu}\,i\pi .
\end{equation}
The above expressions coincide with the $\beta_L\to\infty$, $\beta_R^{-1}=r_0/\pi\ell$ limit of \eqref{eq:btzSEH} and \eqref{eq:btzSanom}.

Now, the holographic TEE at zero temperature with a conserved angular
momentum can be obtained by combining the above expressions followed by \eqref{eq:centralcharges} as
\begin{equation}
S^T_{A,\sigma}
=\frac{c_L}{6}\log\frac{T}{\epsilon}
+\frac{c_R}{6}\log\!\left[\frac{\ell}{r_0\epsilon}\,\sinh\frac{r_0 T}{\ell}\right]
+\frac{c_\sigma}{6}\,i\pi .
\label{eq:extremalresult}
\end{equation}
This agrees with the field-theory result \eqref{eq:ft-TEE-ext} and for $\sigma=-1$ with the piecewise result of \cite{Chu:2025sjv}. In the limit $r_0T/\ell\ll1$ it reduces to \eqref{eq:mainvac} and for $c_L=c_R$ the dependence on $\sigma$ disappears.

We would like to remark that the holographic TEE obtained in this section is determined by a single, continuous complex geodesic whose data are specified at the real asymptotic boundary, together with the orientation of the worldline. This contrasts with the piecewise construction of \cite{Chu:2025sjv}, in which the extremal curve is decomposed into spacelike and timelike geodesic segments and an additional normal frame matching prescription is required at their junctions. The complex geodesic formulation therefore provides a more unified and technically economical description of holographic TEE.

\subsection{Relation to KSW-admissible contours}
\label{sec:ksw}

The complexified prescription used above chooses a contour in the complex affine plane and a branch for each $\log(-1)$. It is natural to ask whether the bulk itself selects that contour. For Einstein gravity this was explained in \cite{Guo:2026vkl} which promotes a family of complex extremal surfaces to a candidate complex bulk metric and imposes the Kontsevich-Segal-Witten (KSW) admissibility condition as
\begin{equation}
\Theta[g]\equiv\sum_\gamma\big|\arg\Lambda_\gamma\big|<\pi
\qquad(\Lambda_\gamma=\text{eigenvalues in a real diagonal frame}),
\label{eq:ksw}
\end{equation}
pointwise along the contour. Real Lorentzian sections saturate the bound, $\Theta=\pi$, and are admitted as its boundary, so that \eqref{eq:CAdS} below is admissible in this sense. For a pure timelike interval the family built from \eqref{eq:complexgeo} induces the single complex metric
\begin{equation}
\frac{\mathrm{d} s^2_{\mathbb C}}{\ell^2}=\mathrm{d}\lambda^2+\cosh^2\lambda\,\mathrm{d}\xi^2-\sinh^2\lambda\,\mathrm{d}\eta^2,
\end{equation}
and Poincar\'e AdS$_3$, global AdS$_3$ and nonrotating BTZ all reduce to the same local complex metric. Upon writing $\lambda=g+i f$, the admissibility condition \eqref{eq:ksw} becomes
\begin{equation}
\Big|\frac{f'}{g'}\Big|\le\Big|\frac{\sin 2f}{\sinh 2g}\Big|,
\end{equation}
and a monotonicity argument then forces a unique admissible contour \cite{Guo:2026vkl}
\begin{equation}
C_{\rm AdS}:\quad
\lambda=\begin{cases}
g-\tfrac{i\pi}{2}, & g<0,\\
i f, & g=0,\ f\in(-\tfrac{\pi}{2},\tfrac{\pi}{2}),\\
g+\tfrac{i\pi}{2}, & g>0,
\end{cases}
\label{eq:CAdS}
\end{equation}
with the spacelike case forced onto the real section $f=0$. This contour has the same endpoints as \eqref{eq:complexgeo} and is homotopic to it in the $\lambda$-plane, so it gives the same length. Its two real segments, $\lambda=g\mp\frac{i\pi}{2}$, are precisely the spacelike geodesics $t^2-u^2=T^2/4$ of the piecewise construction, and the corners $\lambda=\pm\frac{i\pi}{2}$ are the points $u\to\infty$ where those geodesics are joined in \cite{Chu:2025sjv}. The middle segment lies in the section with $t$ and $u$ imaginary rather than on the real timelike geodesic of \cite{Chu:2025sjv}. The branch of the length is therefore selected by the bulk admissibility condition rather than imposed by hand. Moreover, the regularized result of length \cite{Guo:2026vkl}
\begin{equation}
    L_\gamma=2\ell\log\frac{T}{\epsilon}+i\pi\ell
\end{equation}
reproduces our Einstein-Hilbert contribution $S_{\rm EH}$ in the non-anomalous limit $c_L=c_R$. Reversing the orientation of $C_{\rm AdS}$ i.e. taking its mirror image $f\to-f$ or equivalently changing the sign of the boundary $i\epsilon$ prescription, maps $S^T_{A,\sigma}$ to its complex conjugate for either $\sigma$. The $i\epsilon$ choice entering the TEE is therefore encoded geometrically in the orientation of the bulk contour.

The KSW condition constrains only the phases of the metric eigenvalues and is blind to the frame twist which enters through $\eta$ rather than through $\sqrt{-g}$. It therefore fixes the branch of the length, $\frac{c_L+c_R}{12}i\pi$, but not that of the twist. As shown in \cref{sec:ft} and \ref{sec:boosted}, the latter is fixed by the orientation of the interval, and the two orientations give $\pm\frac{c_L-c_R}{12}i\pi$. Whether the replica path integral of the anomalous theory selects one orientation, or whether both define legitimate pseudo-entropies, is a question that KSW admissibility alone cannot settle.

We emphasize two further points. First, the BTZ contour admits an AdS-Rindler description only locally. Globally its timelike segment crosses the outer horizon and for $r_->0$, the inner horizon as well. We have not analysed this global extension and the results \eqref{eq:btzresult} and \eqref{eq:extremalresult} should be understood as local statements. A complete treatment should check that the framing of the twist along the full contour is the one fixed by the orientation, $\mathrm{Im}(\eta_f-\eta_i)=-\sigma\pi$.
Second, the KSW obstruction for timelike strips in AdS$_4$ shows that admissibility is not guaranteed even for the metric contribution \cite{Guo:2026vkl}. Any extension to higher dimensions or to backgrounds that are not locally AdS$_3$ such as warped AdS$_3$, requires an independent analysis of both the metric and the anomalous sectors.

\subsection{Vectors and framing in complexified geometry}
\label{sec:framing}

The computations in preceding sections use a frame $(v,q,\tilde q)$ and its parallel transport along a curve with complex coordinates. Here, we briefly state in what sense these notions are well defined. Let $M$ be the real spacetime with real-analytic metric and $M_{\mathbb C}$ its complexification. The vectors are holomorphic sections of the tangent bundle $T^{1,0}M_{\mathbb C}$, i.e. $\mathbb{C}$-linear combinations of $\partial_\mu$ and the metric continues to a holomorphic symmetric bilinear (not Hermitian) form $g:T\otimes T\to\mathbb{C}$. For the metrics considered here the components are rational in the coordinates, so the continuation is unambiguous away from $u=0$ and the horizon and singularity loci. The Christoffel symbols and curvature are rational in $g$ and $\partial g$, and $\nabla g=0$ is an algebraic identity. The geodesic and parallel-transport equations are holomorphic ODEs: their solutions exist and are unique locally, and they extend by analytic continuation along paths in the complex affine-parameter plane, depending on the path only through its homotopy class relative to the singular points of the connection. The conditions \eqref{eq:frameconds} are holomorphic and preserved by transport. The frame group is $SO(3,\mathbb{C})$, and the residual group fixing $v$ is $SO(2,\mathbb{C})\simeq SO(1,1;\mathbb{C})$, of which the boost $SO(1,1;\mathbb{R})$ and the rotation $SO(2;\mathbb{R})$ of the piecewise construction are the two real forms. Accordingly a single complex parameter $\eta$ in \eqref{eq:twist} describes both, and the reduction of the twist integral to $\eta_f-\eta_i$ goes through unchanged.

Three features of real Lorentzian geometry are lost. First, causal character: $g(v,v)\in\mathbb{C}$, so ``timelike'' and ``spacelike'' have no meaning at generic points of a complex geodesic. Second, positivity: there is no norm and no minimization, so ``minimal'' degenerates to ``extremal'', and the choice among several complex extremal surfaces must be supplied by a physical criterion. Third, a preferred real slice: the decomposition into spacelike segment and a timelike segment is a property of a particular contour, not an invariant, and the timelike segment lies in a different real section of $M_{\mathbb C}$ (with $t$ and $u$ imaginary) rather than in the original spacetime. Consistently, the piecewise curve is not an extremum of the real Lorentzian length functional, its conserved momentum $p=-t'/u^2$ would have to jump by a factor of $i$ at the junctions whereas on the complex geodesic it is conserved throughout.

What survives beyond the local holomorphic data are two discrete choices. The first is the branch of the square root in the unit tangent $v=\dot X/\sqrt{g(\dot X,\dot X)}$, which fixes the sign of the imaginary part of the length, it is fixed by the boundary $i\epsilon$ or equivalently by the orientation of the KSW contour. The second concerns the twist. On a real spacelike geodesic the residual frame group is $SO(1,1;\mathbb R)\simeq\mathbb R$, which is simply connected, so $\eta_f-\eta_i$ in \eqref{eq:twist} is determined by the endpoint data. On a complex geodesic it is $SO(2,\mathbb C)\simeq\mathbb C^*$, with $\pi_1=\mathbb Z$, and the endpoint data fix $\eta_f-\eta_i$ only modulo $2\pi i$. 
A shift $\eta\to\eta+2\pi i$ changes $S_{\rm anom}$ by $\frac{2\pi i}{4G^{(3)}_N\mu}=\frac{c_R-c_L}{6}\,i\pi$, which is exactly the difference between the two orientations in \eqref{eq:ft-TEE-boxed}. The ambiguity has a simple geometric origin. At the tip $\lambda=0$ of \eqref{eq:complexgeo} the tangent is $v^\mu=(T/2,0,0)$ which is parallel to $\partial_t$, so the normal frame induced by the boundary time direction degenerates there and in passing the tip the frame turns by half a revolution, in one sense or the other. For a spacelike interval $\partial_t$ is never tangent to the geodesic and no such choice arises. The orientation of the worldline selects the sense of the half-turn and the relation $\mathrm{Im}(\eta_f-\eta_i)=-\sigma\pi$ of \eqref{eq:twist-boosted} is the bulk counterpart of the boundary statement in \cref{sec:ft}. Finally, the holomorphic formalism does not by itself guarantee that the continued saddle contributes to the continued replica path integral, for the metric sector this is supported by the KSW analysis of \cref{sec:ksw}.

\section{Holographic TEE in Chern-Simons formulation}
\label{sec:cs}

We now obtain the holographic TEE in presence of gravitational anomaly alternatively using the Chern-Simons (CS) formulation of TMG and the Wilson-line construction of the entanglement functional \cite{Castro:2014tta}. Since the flat-connection equations are first order, the multiple-saddle ambiguity of the Mathisson-Papapetrou-Dixon (MPD) description of the probe disappears in this formulation. 
The gravitational anomaly is encoded directly in the unequal Chern-Simons levels $k_L\neq k_R$. Moreover, the branch ambiguity arising in the metric formulation is reinterpreted as the analytic continuation of 
a group holonomy from $SL(2,\mathbb{R})$ to a different real form inside $SL(2,\mathbb{C})$, and its dependence on the orientation of the interval becomes the dependence on the orientation of the Wilson line.

\subsection{TMG as two Chern-Simons connections with a constraint}

Write the dreibein $e^a$ and dualized spin connection $\omega^a=-\tfrac12\epsilon^{abc}\omega_{bc}$ in the two combinations \cite{Deser:1991qk,Carlip:2008qh,Chen:2025xlo}
\begin{equation}
A_L=\Big(\omega^a+\tfrac1\ell e^a\Big)J_a,\qquad
A_R=\Big(\omega^a-\tfrac1\ell e^a\Big)\bar J_a,
\qquad A_{L,R}\in\mathfrak{sl}(2,\mathbb{R}),
\label{eq:ALR}
\end{equation}
where $J_a$ and $\bar J_a$ are two copies of $\mathfrak{sl}(2,\mathbb R)$ generators, written below in the basis $J_0,J_{\pm1}$. Up to the massive constraint, the action for TMG action \eq{eq:TMG-action} is a difference of two CS actions at unequal levels which is given by\cite{Chen:2011yx}
\begin{equation}
S_{\rm TMG}=\Big(1-\tfrac{1}{\mu\ell}\Big)S_{\rm CS}[A_L]-\Big(1+\tfrac{1}{\mu\ell}\Big)S_{\rm CS}[A_R]
-\frac{k}{4\pi\mu\ell}\!\int\mathrm{Tr}\!\big(\beta\wedge(F_L-F_R)\big),
\end{equation}
\begin{equation}
S_{\rm CS}[A]=\frac{k}{4\pi}\int\mathrm{Tr}\Big(A\wedge d A+\tfrac23 A\wedge A\wedge A\Big),
\qquad k=\frac{\ell}{4G^{(3)}_N},
\end{equation}
where $\beta$ is a one-form Lagrange multiplier enforcing torsion free condition $F_L=F_R$. The left and right levels carry the anomaly as
\begin{equation}
k_L=\frac{c_L}{6}=k\Big(1-\frac{1}{\mu\ell}\Big),
\qquad
k_R=\frac{c_R}{6}=k\Big(1+\frac{1}{\mu\ell}\Big),
\qquad
k_R-k_L=\frac{2k}{\mu\ell}=\frac{1}{2G^{(3)}_N\mu}.
\label{eq:levels}
\end{equation}
The gravitational anomaly of the dual CFT is then the case $k_L\neq k_R$. For our analysis, we restrict to locally AdS$_3$ solutions, i.e. flat connections $F_{L,R}=0$, for which the constraint is trivially solved and $A_{L,R}$ are pure gauge locally.  We use the generators in fundamental representation following the convention of \cite{Castro:2014tta}
\begin{equation}
J_0=\tfrac12\begin{pmatrix}1&0\\0&-1\end{pmatrix},\qquad
J_1=\begin{pmatrix}0&0\\-1&0\end{pmatrix},\qquad
J_{-1}=\begin{pmatrix}0&1\\0&0\end{pmatrix},
\label{eq:generators}
\end{equation}
which satisfy $[J_m,J_n]=(m-n)J_{m+n}$. In the following $\mathrm{tr}_f$ denotes the trace in this two-dimensional representation and the same matrices are used for the right generators.

\subsection{Wilson line}\label{sec:wilson}

As shown in \cite{Ammon:2013hba,Castro:2014tta}, the entanglement functional is a Wilson line in a representation $\mathcal R$ of $SO(2,2)\simeq SL(2,\mathbb{R})_L\times SL(2,\mathbb{R})_R$ as
\begin{equation}
W_{\mathcal R}(C)=\mathrm{tr}_{\mathcal R}\,\mathcal{P}\!\exp\!\int_C \mathcal A,
\qquad \mathcal A=A_L\oplus A_R ,
\label{eq:WR}
\end{equation}
in the infinite-dimensional highest-weight representation built on $|h,\bar h\rangle$ with $J_1|h,\bar h\rangle=0$, $J_0|h,\bar h\rangle=h|h,\bar h\rangle$, and similarly for the right generators. The mass and spin of the probe are
\begin{equation}
\ell\,\mathfrak m=h+\bar h,\qquad \mathfrak s=\bar h-h,
\qquad
h=\frac{c_L}{24}\Big(n-\tfrac1n\Big),\ \ \bar h=\frac{c_R}{24}\Big(n-\tfrac1n\Big).
\label{eq:hhbar}
\end{equation}
To first order in $n-1$, $\ell\mathfrak m\simeq(n-1)\frac{\ell}{4G^{(3)}_N}$ and $\mathfrak s\simeq\frac{n-1}{4G^{(3)}_N\mu}$, matching the two coefficients in \eqref{eq:onshell}. Since $\mathcal A=A_L\oplus A_R$, the Wilson line factorizes as
\begin{equation}
W_{\mathcal R}(C)=W_L(C)\,W_R(C),
\qquad
W_{L}(C)=\mathrm{tr}_{\mathcal R}\,\mathcal{P}\!\exp\!\int_C A_L,\ \ \text{etc.}
\label{eq:factor}
\end{equation}
The trace is replaced by a path integral over an auxiliary quantum system on the curve $C$ by
\begin{equation}
W_{L,R}(C)=\int\mathcal{D}U_{L,R}\,\mathrm{e}^{-S_{L,R}},
\end{equation}
with $U_{L,R}\in SL(2,\mathbb{R})$, conjugate momenta $P_{L,R}\in\mathfrak{sl}(2,\mathbb{R})$ and actions
\begin{align}
S_L&=\int_C ds\,\Big[\mathrm{Tr}\big(P_L\,D_sU_L\,U_L^{-1}\big)+\lambda_L\big(\mathrm{Tr}P_L^2-c_2\big)\Big],\qquad D_sU_L=\partial_sU_L+A_LU_L,
\label{eq:SL}\\
S_R&=\int_C ds\,\Big[\mathrm{Tr}\big(P_R\,U_R^{-1}D_sU_R\big)+\lambda_R\big(\mathrm{Tr}P_R^2-\bar c_2\big)\Big],\qquad D_sU_R=\partial_sU_R-U_RA_R,
\label{eq:SR}
\end{align}
where $A_{L,R}$ denote the pullbacks to $C$. They are invariant under $U_L\to LU_L$, $A_L\to L(A_L+d)L^{-1}$ and $U_R\to U_RR$, $A_R\to R^{-1}(A_R+d)R$. The representation enters through the quadratic Casimirs $c_2,\bar c_2$. Requiring the backreaction of the probe to produce the replica conical defect fixes them to the following \cite{Castro:2014tta}
\begin{equation}
\sqrt{2c_2}=\frac{c_L}{6}(n-1)+O\big((n-1)^2\big),\qquad
\sqrt{2\bar c_2}=\frac{c_R}{6}(n-1)+O\big((n-1)^2\big),
\label{eq:casimirs}
\end{equation}
so that $\sqrt{2c_2}\simeq2h$ and $\sqrt{2\bar c_2}\simeq2\bar h$ at leading order.

For locally AdS$_3$ backgrounds we use the gauge following \cite{Castro:2014tta}
\begin{equation}
A_L=L\,dL^{-1},\quad L=\mathrm{e}^{-\rho J_0}\exp\!\Big(-\!\int^x\! a\Big),\qquad
A_R=R^{-1}dR,\quad R=\exp\!\Big(\int^x\!\bar a\Big)\,\mathrm{e}^{-\rho J_0},
\label{eq:LR}
\end{equation}
where $\rho$ is the Fefferman-Graham radial coordinate and $a$, $\bar a$ are the boundary connections. For a real background $L$ and $R$ are real $SL(2,\mathbb{R})$ elements. The saddle-point solutions are
\begin{equation}
\begin{aligned}
U_L(s)&=L\big(X(s)\big)\,\mathrm{e}^{-\alpha(s)P_{L,0}}u_0,\qquad &\mathrm{Tr}P_{L,0}^2&=c_2,\\
U_R(s)&=u_0\,\mathrm{e}^{-\bar\alpha(s)P_{R,0}}R\big(X(s)\big),\qquad &\mathrm{Tr}P_{R,0}^2&=\bar c_2,
\end{aligned}
\label{eq:Usol}
\end{equation}
with on-shell actions $S_L=-c_2\Delta\alpha$ and $S_R=-\bar c_2\Delta\bar\alpha$, where $\Delta\alpha=\alpha(s_f)-\alpha(0)$. The rapidities are fixed by the boundary data through traces in the fundamental representation as
\begin{equation}
\begin{aligned}
2\cosh\!\Big(\sqrt{\tfrac{c_2}{2}}\,\Delta\alpha\Big)&=\mathrm{tr}_f\big(L(X_1)L(X_2)^{-1}U_L(s_f)U_L(0)^{-1}\big),\\
2\cosh\!\Big(\sqrt{\tfrac{\bar c_2}{2}}\,\Delta\bar\alpha\Big)&=\mathrm{tr}_f\big(R(X_1)^{-1}R(X_2)U_R(s_f)^{-1}U_R(0)\big),
\end{aligned}
\label{eq:traceLR}
\end{equation}
with $X_1=X(0)$ and $X_2=X(s_f)$ the two endpoints. It was noted in \cite{Castro:2014tta} that an appropriate choice of the boundary values of $U_{L,R}$ reproduces the metric results. One can write the explicit choice as
\begin{equation}
U_L(s_f)U_L(0)^{-1}=\mathrm{e}^{-J_{-1}},\qquad U_R(s_f)^{-1}U_R(0)=\mathrm{e}^{J_1},
\label{eq:probebc}
\end{equation}
which pair each sector with the generator opposite to the one in its gauge function, so that the traces pick up the boundary separations (see \eqref{eq:traceVac} below). Now define $M_L$ and $M_R$ by
\begin{equation}
M_L+M_L^{-1}=\mathrm{tr}_f\big(L(X_1)L(X_2)^{-1}\mathrm{e}^{-J_{-1}}\big),\qquad
M_R+M_R^{-1}=\mathrm{tr}_f\big(R(X_1)^{-1}R(X_2)\,\mathrm{e}^{J_1}\big),
\label{eq:Mdef}
\end{equation}
and taking the root with $|M_{L,R}|\ge1$ i.e. $\sqrt{c_2/2}\,\Delta\alpha=-\log M_L$ and $\sqrt{\bar c_2/2}\,\Delta\bar\alpha=-\log M_R$, the roots that make the entropy of a spacelike interval positive. Then $S_L=\sqrt{2c_2}\log M_L$ and $S_R=\sqrt{2\bar c_2}\log M_R$. In the semiclassical limit $W_{L,R}\simeq\mathrm{e}^{-S_{L,R}}$ and using \eqref{eq:casimirs}, the holographic entanglement entropy can be written as \cite{Castro:2014tta}
\begin{equation}
S=\lim_{n\to1}\frac{1}{1-n}\log\big(W_LW_R\big)
=\frac{c_L}{6}\log M_L+\frac{c_R}{6}\log M_R.
\label{eq:Smaster}
\end{equation}
The anomaly is the unequal weighting of the two chiral sectors by $c_L/6$ and $c_R/6$. In the non-anomalous case $c_L=c_R$ it reduces to $\frac{c}{6}\log(M_LM_R)$. We now evaluate $M_L$ and $M_R$ for timelike intervals.

\subsection{Complex saddle}
\label{sec:cs-direct}

For a timelike interval no real geodesic joins the two boundary points yet the Wilson line is still well defined. The endpoints are real boundary points, $L(X_{1,2})$ and $R(X_{1,2})$ are real $SL(2,\mathbb{R})$ elements and the traces in \eqref{eq:Mdef} are real. The bulk connection is never complexified. What becomes complex is the saddle of the auxiliary probe integral where the rapidities $\Delta\alpha,\Delta\bar\alpha$ in \eqref{eq:Usol} are complex, and the probe trajectory is the complex geodesic \eqref{eq:complexgeo}, along which $L(X(\lambda))$ and $R(X(\lambda))$ run through $SL(2,\mathbb{C})$ in the interior. The curve $C$ may be any curve in the homotopy class. The complex saddle enters only by fixing the branches of the two invariants, which for timelike endpoints are evaluated directly without introducing a spacelike interval.

\paragraph{Traces at the timelike endpoints:}
In Poincar\'e AdS$_3$ the gauge functions \eqref{eq:LR} are given by
\begin{equation}
L=\mathrm{e}^{-\rho J_0}\mathrm{e}^{-J_1z},\qquad R=\mathrm{e}^{-J_{-1}\bar z}\mathrm{e}^{-\rho J_0},
\label{eq:LRvac}
\end{equation}
so that $A_L=J_0\,d\rho+\mathrm{e}^{\rho}J_1\,d z$ and $A_R=-J_0\,d\rho-\mathrm{e}^{\rho}J_{-1}\,d\bar z$. With the endpoints at the cutoff $\mathrm{e}^{\rho}=1/\epsilon$, the traces \eqref{eq:Mdef} can be obtained as
\begin{equation}
M_L+M_L^{-1}=2+\frac{z_{21}}{\epsilon},\qquad M_R+M_R^{-1}=2+\frac{\bar z_{21}}{\epsilon},
\label{eq:traceVac}
\end{equation}
with $z=x-t$, $\bar z=x+t$ and $z_{21}=z_2-z_1$. For the pure timelike interval the endpoints are $(t,x)=(\mp T/2,0)$, so $z_{21}=-T$ and $\bar z_{21}=+T$, and to leading order in $\epsilon$
\begin{equation}
M_L=-\frac{T}{\epsilon},\qquad
M_R=\frac{T}{\epsilon}.
\label{eq:Mtimelike}
\end{equation}
For $X_2$ later than $X_1$, $z_{21}<0<\bar z_{21}$, and the left trace lies below $-2$, so $\cosh(\sqrt{c_2/2}\,\Delta\alpha)<-1$ and the left rapidity is complex. Reversing the orientation of $C$ flips the signs of both separations and moves the negative trace to the right sector. In either case one rapidity is complex, which is the group-theoretic signature of timelike separation, while for a spacelike interval both traces exceed $2$.

\paragraph{Mass-spin form of the Wilson line.}
The probe enters only through its mass and spin which can be observed from \eqref{eq:hhbar}. Accordingly we can write \eqref{eq:Smaster} in the equivalent form as
\begin{equation}
S=\frac{c_L+c_R}{12}\,\log\big(M_LM_R\big)+\frac{c_R-c_L}{12}\,\log\frac{M_R}{M_L},
\label{eq:Smassspin}
\end{equation}
whose coefficients are the replica limits of the mass and spin or equivalently $(\sqrt{2\bar c_2}\pm\sqrt{2 c_2})/2(n-1)$ by \eqref{eq:casimirs}. The first invariant term, $\log(M_LM_R)$ is the length of the geodesic. The second term,  $\log(M_R/M_L)$ is the frame twist. These are the Chern-Simons counterparts of the two separate terms $S_{\rm EH}$ and $S_{\rm anom}$ of \cref{sec:holo}. On the continued branch $\log(M_LM_R)=\log M_L+\log M_R$ and $\log(M_R/M_L)=\log M_R-\log M_L$ hold exactly, so \eqref{eq:Smassspin} is identical to \eqref{eq:Smaster}. Taking the principal value of the product and the ratio separately breaks the second identity by $2\pi i$. For the timelike endpoints \eqref{eq:Mtimelike}, we have
\begin{equation}
M_LM_R=-\frac{T^2}{\epsilon^2},\qquad \frac{M_R}{M_L}=-1 .
\end{equation}
Both invariants are negative and exchanging the endpoints leaves their values unchanged. Their continued logarithms are not unchanged: with the prescription of \cref{sec:ft}, $\log M_L$ acquires $+i\pi$ for $X_2$ later than $X_1$, so $\log(M_LM_R)\to\log\frac{T^2}{\epsilon^2}+i\pi$ and $\log(M_R/M_L)\to-\sigma\,i\pi$, with $\sigma=+1$ for this orientation of $C$. We obtain the following
\begin{equation}
S^T_{A,\sigma}
=\frac{c_L+c_R}{12}\Big(\log\frac{T^2}{\epsilon^2}+i\pi\Big)-\sigma\,\frac{c_R-c_L}{12}\,i\pi
=\frac{c_L+c_R}{6}\log\frac{T}{\epsilon}+\frac{c_\sigma}{6}\,i\pi .
\label{eq:cs-direct}
\end{equation}
We see that the above expression \eq{eq:cs-direct} matches \eqref{eq:mainvac} obtained via the complexified prescription. We further note that evaluating the traces at the timelike endpoints is itself an analytic continuation of the formula \eqref{eq:Smaster}, even though no spacelike interval is introduced. The traces \eqref{eq:traceVac} are exact for any endpoints but three steps in passing from them to \eqref{eq:Smaster} rely on the spacelike case:
\begin{itemize}
\item The evaluation of $W_{L,R}\simeq\mathrm{e}^{-S_{L,R}}$ assumes a saddle on the original cycle $U_{L,R}\in SL(2,\mathbb{R})$, which exists only for traces $\ge2$, whereas for timelike endpoints the saddle is complex.
\item The root $|M_{L,R}|\ge1$ was fixed by positivity of the spacelike entanglement entropy and $\log M$ is defined for $M>0$, whereas $M_L<0$ here.
\item The identification $\mathrm{Tr}\rho^n=W_{\mathcal R}$ and the Casimirs \eqref{eq:casimirs} are derived for spacelike R\'enyi entropies whereas for a timelike interval $\mathrm{Tr}\rho^n$ is a pseudo-R\'enyi quantity.
\end{itemize}
This provides the Chern-Simons counterpart of the analytic continuation employed in the dual field theory computation.

\subsection{Rotating BTZ}
\label{sec:cs-btz}

We now evaluate the Wilson line directly at the endpoints of a pure timelike interval in the rotating BTZ background similar to the preceding \cref{sec:cs-direct}.
For the planar rotating BTZ brane the boundary connections in \eqref{eq:LR} are constant and of the following form \cite{Castro:2014tta}
\begin{equation}
a=\Big(J_1-\frac{\pi^2}{\beta_L^2}J_{-1}\Big)\mathrm{d}z,\qquad
\bar a=-\Big(J_{-1}-\frac{\pi^2}{\beta_R^2}J_1\Big)\mathrm{d}\bar z,
\label{eq:abtz}
\end{equation}
with boundary light-cone coordinates $z=\phi-\tau$ and $\bar z=\phi+\tau$. These reduce to the vacuum gauge functions \eqref{eq:LRvac} as $\beta_{L,R}\to\infty$. In the fundamental representation
\begin{equation}
\Big(J_1-\frac{\pi^2}{\beta_L^2}J_{-1}\Big)^2=\frac{\pi^2}{\beta_L^2}\,\mathbf{1},\qquad
\Big(J_{-1}-\frac{\pi^2}{\beta_R^2}J_1\Big)^2=\frac{\pi^2}{\beta_R^2}\,\mathbf{1},
\end{equation}
so that the exponentials in $L$ and $R$ are elementary and the traces \eqref{eq:Mdef} are obtained as
\begin{equation}
M_L+M_L^{-1}=2\cosh\frac{\pi z_{21}}{\beta_L}+\frac{\beta_L}{\pi\epsilon}\sinh\frac{\pi z_{21}}{\beta_L},\qquad
M_R+M_R^{-1}=2\cosh\frac{\pi\bar z_{21}}{\beta_R}+\frac{\beta_R}{\pi\epsilon}\sinh\frac{\pi\bar z_{21}}{\beta_R},
\label{eq:traceBTZ}
\end{equation}
with the endpoints at the cutoff $\mathrm{e}^{\rho}=1/\epsilon$.
The pure timelike interval has endpoints $(\tau,\phi)=(\mp T/2,0)$, so
\begin{equation}
z_{21}=-T,\qquad \bar z_{21}=T .
\end{equation}
Using these in \eqref{eq:traceBTZ}, we obtain
\begin{equation}
M_L+M_L^{-1}=2\cosh\frac{\pi T}{\beta_L}-\frac{\beta_L}{\pi\epsilon}\sinh\frac{\pi T}{\beta_L},\qquad
M_R+M_R^{-1}=2\cosh\frac{\pi T}{\beta_R}+\frac{\beta_R}{\pi\epsilon}\sinh\frac{\pi T}{\beta_R}.
\end{equation}
For $\epsilon\to0$, with $X_2$ later than $X_1$, the left trace lies below $-2$ and the right trace above $2$, so that to leading order
\begin{equation}
M_L=-\frac{\beta_L}{\pi\epsilon}\sinh\frac{\pi T}{\beta_L},\qquad
M_R=\frac{\beta_R}{\pi\epsilon}\sinh\frac{\pi T}{\beta_R}.
\label{eq:Mbtz}
\end{equation}
Similar to the vacuum case \eqref{eq:Mtimelike}, the left element is negative, $\cosh(\sqrt{c_2/2}\,\Delta\alpha)<-1$ and the left rapidity of the probe saddle is complex.
The two invariants entering the mass-spin form \eqref{eq:Smassspin} are now given by
\begin{equation}
M_LM_R=-\frac{\beta_L\beta_R}{\pi^2\epsilon^2}\,\sinh\frac{\pi T}{\beta_L}\,\sinh\frac{\pi T}{\beta_R},\qquad
\frac{M_R}{M_L}=-\,\frac{\beta_R\sinh\frac{\pi T}{\beta_R}}{\beta_L\sinh\frac{\pi T}{\beta_L}} .
\label{eq:invBTZ}
\end{equation}
As in the vacuum, both are negative and their values do not depend on the labelling of the endpoints while the branch of $\log(M_R/M_L)$ follows the orientation of $C$. They coincide with the quantities of the metric computation of \cref{sec:btz}. Upon utilizing \eq{eq:invBTZ} in \eqref{eq:Smassspin}, the holographic TEE at finite temperature with angular potential can be obtained as
\begin{align}
S^T_{A,\sigma}=\frac{c_L}{6}\log\!\Big[\frac{\beta_L}{\pi\epsilon}\sinh\frac{\pi T}{\beta_L}\Big]
+\frac{c_R}{6}\log\!\Big[\frac{\beta_R}{\pi\epsilon}\sinh\frac{\pi T}{\beta_R}\Big]
+\frac{c_\sigma}{6}\,i\pi .
\label{eq:cs-btz}
\end{align}
We observe that the above result matches with \eqref{eq:btzresult} which is obtained from complexified prescription.

\paragraph{Extremal limit:}
For the extremal black hole, $\beta_L\to\infty$ and $\beta_R^{-1}=r_0/\pi\ell$, the left connection reduces to $a=J_1\,\mathrm{d}z$ and the left trace to the vacuum one, $M_L+M_L^{-1}=2+z_{21}/\epsilon=2-T/\epsilon$. Hence we have
\begin{equation}
M_L=-\frac{T}{\epsilon},\qquad M_R=\frac{\ell}{r_0\epsilon}\sinh\frac{r_0T}{\ell},\qquad
M_LM_R=-\frac{\ell T}{r_0\epsilon^2}\sinh\frac{r_0T}{\ell},\qquad
\frac{M_R}{M_L}=-\frac{\ell}{r_0T}\sinh\frac{r_0T}{\ell},
\end{equation}
which are the arguments of $S_{\rm EH}$ and $S_{\rm anom}$ in \cref{sec:extremal}. With the same branch assignments \eqref{eq:Smassspin} gives \eqref{eq:extremalresult} directly.

Having computed the holographic TEE, some comments are in order. In the Chern-Simons formulation two different objects could be taken off the real domain: the flat connections $A_{L,R}$, by promoting them to $\mathfrak{sl}(2,\mathbb{C})$ and the saddle \eqref{eq:Usol} of the auxiliary probe. Only the second is needed. A timelike boundary interval does not change the background. On real Poincar\'e AdS$_3$ or real BTZ the connections, the gauge functions $L,R$ and the Casimirs \eqref{eq:casimirs} are real, and so are the traces \eqref{eq:Mdef} evaluated at the real endpoints. What changes is the value of one of the traces. For a spacelike interval both traces exceed $2$ and the rapidities are real, whereas for a timelike interval the trace of the sector whose light-cone separation is negative falls below $-2$ by \eqref{eq:traceVac} and \eqref{eq:traceBTZ}, this is the left sector for $X_2$ later than $X_1$. Then $\cosh(\sqrt{c_2/2}\,\Delta\alpha)<-1$ and $\Delta\alpha$ is complex: the probe saddle leaves $SL(2,\mathbb{R})$ and becomes the complex geodesic while the bulk connection stays real. This is the Chern-Simons form of the statement that TEE is complex.
The two branches it requires are fixed as in the boundary CFT, the length branch by the $i\epsilon$ prescription and the twist branch by the orientation of the Wilson line. \Cref{sec:csreal} shows this without reference to the complex saddle.

This is a structural advantage over the metric formulation, where a single extremal carrier of the TEE requires a complex section of the geometry with tip $u_t=iT/2$. In the Chern-Simons formulation the complexity is confined to the probe saddle and the endpoint data, and complexifying the connection is at most a reorganization. The same holds across all the routes considered here. The CFT continuation uses only endpoint data, the piecewise construction uses a real Lorentzian bulk with one timelike segment whose imaginary proper length supplies the $i\pi$, the complexified metric construction uses a single complex geodesic, and the Wilson line keeps the connection real and lets only the probe saddle become complex. All four give the same TEE. The complexified TMG in the sense of a complex metric or a complex connection is therefore sufficient but not necessary and the unavoidable ingredients are the two branch data at timelike separation i.e. the $i\epsilon$ prescription for the length and the orientation for the twist.

\subsection{Branches from the boundary prescription}
\label{sec:csreal}

The traces at the timelike endpoints are real and at the cutoff they reproduce exactly the chiral factors of the CFT two-point function: $M_L\simeq z_{21}/\epsilon$ and $M_R\simeq\bar z_{21}/\epsilon$ in the vacuum \eq{eq:Mtimelike} and the thermal factors \eqref{eq:Mbtz} in the rotating BTZ. The complex saddle is therefore needed only to fix the branches of the two invariants in \eqref{eq:Smassspin}. They can instead be fixed by the boundary time-ordering prescription $t_{21}\to t_{21}(1-i\delta)$, $\delta\to0^+$, together with the orientation of $C$, keeping every bulk quantity real.

Take $X_2$ later than $X_1$ and continue from a spacelike configuration. Then $z_{21}$ crosses zero from above while $\bar z_{21}$ stays positive, and at $x_{21}=0$
\begin{equation}
\arg M_L=\pi-\delta\,a_L\coth a_L+O(\delta^2),\qquad \arg M_R=-\delta\,a_R\coth a_R+O(\delta^2),\qquad a_{L,R}=\frac{\pi T}{\beta_{L,R}},
\end{equation}
with $a\coth a\to1$ in the vacuum. Hence $\log M_L$ acquires $+i\pi$ and $\log M_R$ stays real. Then $\log(M_LM_R)$ acquires $+i\pi$, in agreement with the branch selected by the complex saddle and by the orientation of the KSW contour, and $\log(M_R/M_L)$ acquires $-i\pi$. This holds in the vacuum and in rotating and extremal BTZ alike. 
For $\beta_L>\beta_R$ the ratio $M_R/M_L$ lies just above the negative real axis, so its principal logarithm is $+i\pi$, this differs from the continued value by $2\pi i$ and is not the analytic continuation. Reversing the orientation of $C$ exchanges the roles of the two sectors and gives $\log(M_R/M_L)\to+i\pi$. With these branches \eqref{eq:Smassspin} gives $\frac{c_L}{6}i\pi$ for $C$ future-directed and $\frac{c_R}{6}i\pi$ for $C$ past-directed, reproducing \eqref{eq:cs-direct} and \eqref{eq:cs-btz} without reference to a complex geometry. The Wilson line thus makes the orientation dependence manifest where the spinning probe is an oriented worldline and its two orientations are the two values of $\sigma$.

\section{Summary and conclusion}
\label{sec:summary}

To summarize, we have studied the timelike entanglement entropy in a CFT$_2$ with a gravitational anomaly, holographically dual to topologically massive gravity in AdS$_3$. Using the complexified-geometry prescription and its Chern-Simons/Wilson-line realization, we reproduced the field-theory results for pure and boosted timelike intervals in Poincar\'e AdS$_3$ and for pure timelike intervals in the non-extremal and extremal rotating BTZ backgrounds. We observed that the real part depends on both central charges and not on the orientation. In contrast, the imaginary part depends on a single central charge, $c_L$ or $c_R$ according to whether the twist operator sits at the later or the earlier endpoint. The difference between the two orientations is fixed by the anomaly alone and persists at finite R\'enyi index.

On the gravity side, the holographic TEE is described by the on-shell action of a massive spinning particle evaluated on a single complex geodesic anchored on the real timelike interval which is the geodesic length plus the twist of a parallel-transported normal frame. This replaces the piecewise construction of spacelike and timelike geodesics. The boost and rotation of the normal frame become two real forms of one complexified residual group where boundary data are needed only at the real boundary and the free parameter of the timelike segment is fixed. 
We then computed the holographic TEE in a CFT$_2$ dual to pure AdS$_3$, rotating non-extremal and extremal BTZ black holes utilizing this prescription.
The holographic results in all these cases match the dual field theory results.
Further, the real segments of the KSW contour reproduce the spacelike geodesics of the piecewise construction. Moreover, the metric sector contour is not imposed independently but is uniquely fixed by the Kontsevich-Segal–Witten (KSW) admissibility criterion in locally AdS$_3$ backgrounds.
In BTZ this identification is local since the global contour crosses the horizons.

The structure of the imaginary part rests on the fact that the holographic functional is a sum of two separately computed terms, the length and the twist, equivalently the mass and spin invariants of the probe. Each produces its own $\log(-1)$, and the two branches are fixed by different data. The length branch is fixed by the boundary time-ordering prescription or equivalently by the orientation of the KSW contour. The twist branch is fixed by the orientation of the interval i.e. by which endpoint carries the twist operator, and in the bulk by the orientation of the spinning worldline. The two orientations give $\frac{c_L}{6}i\pi$ and $\frac{c_R}{6}i\pi$, the principal value prescription of \cite{Chu:2025sjv} reproduces the second for either labelling. In the bulk the ambiguity is a framing ambiguity which appears because the residual frame group of a complex geodesic, $SO(2,\mathbb C)\simeq\mathbb C^*$ is not simply connected, while that of a real spacelike geodesic is.

In the Chern--Simons formulation the anomaly appears as the level asymmetry $k_{L,R}=c_{L,R}/6$, and following \cite{Castro:2014tta} the entanglement entropy is fixed by two exact traces of the gauge functions at the endpoints, one per chirality. For a timelike interval one of the two traces falls below $-2$, so the corresponding probe rapidity and hence the probe saddle becomes complex. This saddle is the complex geodesic while the bulk connection stays real. Which trace becomes negative is set by the orientation of the Wilson line, so the orientation dependence found on the field theory side is built into the probe. Complexifying TMG in the sense of a complex metric or a complex connection is therefore sufficient but not necessary, and both branches can be fixed by the boundary prescription and the orientation alone. Finally, vectors, frames and parallel transport in the complexified geometry are well defined as holomorphic Riemannian geometry while causal character, positivity and a preferred real slice are lost. What survives are the branch of the length, fixed by the $i\epsilon$ prescription, and the framing of the twist which is fixed by the orientation.

Several directions remain open. The most direct is to decide which orientation, if either, is singled out by an operational definition of entanglement in time. A computation of the timelike entanglement entropy for free chiral fermions with $c_L\neq c_R$ along the lines of \cite{Liu:2022ugc}, or of the entanglement in time of \cite{Milekhin:2025ycm} would settle this. A related question is whether the replica path integral of the anomalous theory, or a Picard-Lefschetz analysis of the probe path integral selects one orientation, since the KSW criterion does not. Other directions include the global form of the BTZ contour for $r_->0$ which crosses the inner horizon and its relation to the inner RT surface of \cite{Wen:2024muv}. Further extensions include extension to higher dimensions, to warped AdS$_3$ and to other backgrounds that are not locally AdS$_3$ where KSW admissibility is not guaranteed even in the metric sector. We leave these open issues for future investigations.

\section*{Acknowledgments}
We would like to thank Fabio Ori for helpful discussions. J.Y. was supported by the National Research Foundation of Korea (NRF) grant funded by the Korean government (MSIT) (RS-2022-NR069038, RS-2025-25466315). J.Y. and H.P. were supported by Brain Pool program funded by the Ministry of Science and ICT through the National Research Foundation of Korea (RS-2025-25457100). This work was partly supported by the Institute of Information \& Communications Technology Planning \& Evaluation(IITP)-ITRC(Information Technology Research Center) grant funded by the Korea government(MSIT)(IITP-2026-RS-2026-25519864).

\bibliographystyle{JHEP}

\bibliography{ref} 

\providecommand{\href}[2]{#2}\begingroup\raggedright\begin{thebibliography}{10}

\bibitem{Calabrese:2004eu}
P.~Calabrese and J.L.~Cardy, \emph{{Entanglement entropy and quantum field theory}}, \href{https://doi.org/10.1088/1742-5468/2004/06/P06002}{\emph{J. Stat. Mech.} {\bfseries 0406} (2004) P06002} [\href{https://arxiv.org/abs/hep-th/0405152}{{\ttfamily hep-th/0405152}}].

\bibitem{Calabrese:2009qy}
P.~Calabrese and J.~Cardy, \emph{{Entanglement entropy and conformal field theory}}, \href{https://doi.org/10.1088/1751-8113/42/50/504005}{\emph{J. Phys. A} {\bfseries 42} (2009) 504005} [\href{https://arxiv.org/abs/0905.4013}{{\ttfamily 0905.4013}}].

\bibitem{Ryu:2006bv}
S.~Ryu and T.~Takayanagi, \emph{{Holographic derivation of entanglement entropy from AdS/CFT}}, \href{https://doi.org/10.1103/PhysRevLett.96.181602}{\emph{Phys. Rev. Lett.} {\bfseries 96} (2006) 181602} [\href{https://arxiv.org/abs/hep-th/0603001}{{\ttfamily hep-th/0603001}}].

\bibitem{Ryu:2006ef}
S.~Ryu and T.~Takayanagi, \emph{{Aspects of Holographic Entanglement Entropy}}, \href{https://doi.org/10.1088/1126-6708/2006/08/045}{\emph{JHEP} {\bfseries 08} (2006) 045} [\href{https://arxiv.org/abs/hep-th/0605073}{{\ttfamily hep-th/0605073}}].

\bibitem{Alvarez-Gaume:1983ihn}
L.~Alvarez-Gaume and E.~Witten, \emph{{Gravitational Anomalies}}, \href{https://doi.org/10.1016/0550-3213(84)90066-X}{\emph{Nucl. Phys. B} {\bfseries 234} (1984) 269}.

\bibitem{Deser:1981wh}
S.~Deser, R.~Jackiw and S.~Templeton, \emph{{Topologically Massive Gauge Theories}}, \href{https://doi.org/10.1016/0003-4916(82)90164-6}{\emph{Annals Phys.} {\bfseries 140} (1982) 372}.

\bibitem{Kraus:2005zm}
P.~Kraus and F.~Larsen, \emph{{Holographic gravitational anomalies}}, \href{https://doi.org/10.1088/1126-6708/2006/01/022}{\emph{JHEP} {\bfseries 01} (2006) 022} [\href{https://arxiv.org/abs/hep-th/0508218}{{\ttfamily hep-th/0508218}}].

\bibitem{Castro:2014tta}
A.~Castro, S.~Detournay, N.~Iqbal and E.~Perlmutter, \emph{{Holographic entanglement entropy and gravitational anomalies}}, \href{https://doi.org/10.1007/JHEP07(2014)114}{\emph{JHEP} {\bfseries 07} (2014) 114} [\href{https://arxiv.org/abs/1405.2792}{{\ttfamily 1405.2792}}].

\bibitem{Ammon:2013hba}
M.~Ammon, A.~Castro and N.~Iqbal, \emph{{Wilson Lines and Entanglement Entropy in Higher Spin Gravity}}, \href{https://doi.org/10.1007/JHEP10(2013)110}{\emph{JHEP} {\bfseries 10} (2013) 110} [\href{https://arxiv.org/abs/1306.4338}{{\ttfamily 1306.4338}}].

\bibitem{He:2023cju}
P.-Z.~He and H.-Q.~Zhang, \emph{{Revisit the entanglement entropy with gravitational anomaly}}, \href{https://doi.org/10.1007/JHEP11(2023)142}{\emph{JHEP} {\bfseries 11} (2023) 142} [\href{https://arxiv.org/abs/2305.05892}{{\ttfamily 2305.05892}}].

\bibitem{Basu:2022nds}
D.~Basu, H.~Parihar, V.~Raj and G.~Sengupta, \emph{{Entanglement negativity, reflected entropy, and anomalous gravitation}}, \href{https://doi.org/10.1103/PhysRevD.105.086013}{\emph{Phys. Rev. D} {\bfseries 105} (2022) 086013} [\href{https://arxiv.org/abs/2202.00683}{{\ttfamily 2202.00683}}].

\bibitem{Wen:2022jxr}
Q.~Wen and H.~Zhong, \emph{{Covariant entanglement wedge cross-section, balanced partial entanglement and gravitational anomalies}}, \href{https://doi.org/10.21468/SciPostPhys.13.3.056}{\emph{SciPost Phys.} {\bfseries 13} (2022) 056} [\href{https://arxiv.org/abs/2205.10858}{{\ttfamily 2205.10858}}].

\bibitem{Doi:2022iyj}
K.~Doi, J.~Harper, A.~Mollabashi, T.~Takayanagi and Y.~Taki, \emph{{Pseudoentropy in dS/CFT and Timelike Entanglement Entropy}}, \href{https://doi.org/10.1103/PhysRevLett.130.031601}{\emph{Phys. Rev. Lett.} {\bfseries 130} (2023) 031601} [\href{https://arxiv.org/abs/2210.09457}{{\ttfamily 2210.09457}}].

\bibitem{Doi:2023zaf}
K.~Doi, J.~Harper, A.~Mollabashi, T.~Takayanagi and Y.~Taki, \emph{{Timelike entanglement entropy}}, \href{https://doi.org/10.1007/JHEP05(2023)052}{\emph{JHEP} {\bfseries 05} (2023) 052} [\href{https://arxiv.org/abs/2302.11695}{{\ttfamily 2302.11695}}].

\bibitem{Nakata:2020luh}
Y.~Nakata, T.~Takayanagi, Y.~Taki, K.~Tamaoka and Z.~Wei, \emph{{New holographic generalization of entanglement entropy}}, \href{https://doi.org/10.1103/PhysRevD.103.026005}{\emph{Phys. Rev. D} {\bfseries 103} (2021) 026005} [\href{https://arxiv.org/abs/2005.13801}{{\ttfamily 2005.13801}}].

\bibitem{Mollabashi:2020yie}
A.~Mollabashi, N.~Shiba, T.~Takayanagi, K.~Tamaoka and Z.~Wei, \emph{{Pseudo Entropy in Free Quantum Field Theories}}, \href{https://doi.org/10.1103/PhysRevLett.126.081601}{\emph{Phys. Rev. Lett.} {\bfseries 126} (2021) 081601} [\href{https://arxiv.org/abs/2011.09648}{{\ttfamily 2011.09648}}].

\bibitem{Mollabashi:2021xsd}
A.~Mollabashi, N.~Shiba, T.~Takayanagi, K.~Tamaoka and Z.~Wei, \emph{{Aspects of pseudoentropy in field theories}}, \href{https://doi.org/10.1103/PhysRevResearch.3.033254}{\emph{Phys. Rev. Res.} {\bfseries 3} (2021) 033254} [\href{https://arxiv.org/abs/2106.03118}{{\ttfamily 2106.03118}}].

\bibitem{Milekhin:2025ycm}
A.~Milekhin, Z.~Adamska and J.~Preskill, \emph{{Observable and computable entanglement in time}},  \href{https://arxiv.org/abs/2502.12240}{{\ttfamily 2502.12240}}.

\bibitem{Li:2022tsv}
Z.~Li, Z.-Q.~Xiao and R.-Q.~Yang, \emph{{On holographic time-like entanglement entropy}}, \href{https://doi.org/10.1007/JHEP04(2023)004}{\emph{JHEP} {\bfseries 04} (2023) 004} [\href{https://arxiv.org/abs/2211.14883}{{\ttfamily 2211.14883}}].

\bibitem{Basak:2023otu}
J.K.~Basak, A.~Chakraborty, C.-S.~Chu, D.~Giataganas and H.~Parihar, \emph{{Massless Lifshitz field theory for arbitrary z}}, \href{https://doi.org/10.1007/JHEP05(2024)284}{\emph{JHEP} {\bfseries 05} (2024) 284} [\href{https://arxiv.org/abs/2312.16284}{{\ttfamily 2312.16284}}].

\bibitem{Afrasiar:2024lsi}
M.~Afrasiar, J.K.~Basak and D.~Giataganas, \emph{{Timelike entanglement entropy and phase transitions in non-conformal theories}}, \href{https://doi.org/10.1007/JHEP07(2024)243}{\emph{JHEP} {\bfseries 07} (2024) 243} [\href{https://arxiv.org/abs/2404.01393}{{\ttfamily 2404.01393}}].

\bibitem{Afrasiar:2024ldn}
M.~Afrasiar, J.K.~Basak and D.~Giataganas, \emph{{Holographic Timelike Entanglement Entropy in Non-relativistic Theories}},  \href{https://arxiv.org/abs/2411.18514}{{\ttfamily 2411.18514}}.

\bibitem{Afrasiar:2025eam}
M.~Afrasiar, J.K.~Basak and K.-Y.~Kim, \emph{{Aspects of holographic timelike entanglement entropy in black hole backgrounds}},  \href{https://arxiv.org/abs/2512.21327}{{\ttfamily 2512.21327}}.

\bibitem{Heller:2024whi}
M.P.~Heller, F.~Ori and A.~Serantes, \emph{{Geometric Interpretation of Timelike Entanglement Entropy}}, \href{https://doi.org/10.1103/PhysRevLett.134.131601}{\emph{Phys. Rev. Lett.} {\bfseries 134} (2025) 131601} [\href{https://arxiv.org/abs/2408.15752}{{\ttfamily 2408.15752}}].

\bibitem{Guo:2026vkl}
W.-z.~Guo, \emph{{Selecting Complex Extremal Surfaces with the Kontsevich--Segal--Witten Criterion}},  \href{https://arxiv.org/abs/2607.21030}{{\ttfamily 2607.21030}}.

\bibitem{Kontsevich:2021dmb}
M.~Kontsevich and G.~Segal, \emph{{Wick Rotation and the Positivity of Energy in Quantum Field Theory}}, \href{https://doi.org/10.1093/qmath/haab027}{\emph{Quart. J. Math. Oxford Ser.} {\bfseries 72} (2021) 673} [\href{https://arxiv.org/abs/2105.10161}{{\ttfamily 2105.10161}}].

\bibitem{Witten:2021nzp}
E.~Witten, \emph{{A Note On Complex Spacetime Metrics}},  \href{https://arxiv.org/abs/2111.06514}{{\ttfamily 2111.06514}}.

\bibitem{Wang:2018jva}
P.~Wang, H.~Wu and H.~Yang, \emph{{Fix the dual geometries of $T\bar{T}$ deformed CFT$_2$ and highly excited states of CFT$_2$}}, \href{https://doi.org/10.1140/epjc/s10052-020-08680-7}{\emph{Eur. Phys. J. C} {\bfseries 80} (2020) 1117} [\href{https://arxiv.org/abs/1811.07758}{{\ttfamily 1811.07758}}].

\bibitem{Liu:2022ugc}
B.~Liu, H.~Chen and B.~Lian, \emph{{Entanglement entropy of free fermions in timelike slices}}, \href{https://doi.org/10.1103/PhysRevB.110.144306}{\emph{Phys. Rev. B} {\bfseries 110} (2024) 144306} [\href{https://arxiv.org/abs/2210.03134}{{\ttfamily 2210.03134}}].

\bibitem{Narayan:2022afv}
K.~Narayan, \emph{{de Sitter space, extremal surfaces, and time entanglement}}, \href{https://doi.org/10.1103/PhysRevD.107.126004}{\emph{Phys. Rev. D} {\bfseries 107} (2023) 126004} [\href{https://arxiv.org/abs/2210.12963}{{\ttfamily 2210.12963}}].

\bibitem{Jiang:2023ffu}
X.~Jiang, P.~Wang, H.~Wu and H.~Yang, \emph{{Timelike entanglement entropy and TT\textasciimacron{} deformation}}, \href{https://doi.org/10.1103/PhysRevD.108.046004}{\emph{Phys. Rev. D} {\bfseries 108} (2023) 046004} [\href{https://arxiv.org/abs/2302.13872}{{\ttfamily 2302.13872}}].

\bibitem{Narayan:2023ebn}
K.~Narayan and H.K.~Saini, \emph{{Notes on time entanglement and pseudo-entropy}}, \href{https://doi.org/10.1140/epjc/s10052-024-12855-x}{\emph{Eur. Phys. J. C} {\bfseries 84} (2024) 499} [\href{https://arxiv.org/abs/2303.01307}{{\ttfamily 2303.01307}}].

\bibitem{Chu:2023zah}
C.-S.~Chu and H.~Parihar, \emph{{Time-like entanglement entropy in AdS/BCFT}}, \href{https://doi.org/10.1007/JHEP06(2023)173}{\emph{JHEP} {\bfseries 06} (2023) 173} [\href{https://arxiv.org/abs/2304.10907}{{\ttfamily 2304.10907}}].

\bibitem{Jiang:2023loq}
X.~Jiang, P.~Wang, H.~Wu and H.~Yang, \emph{{Timelike entanglement entropy in dS$_{3}$/CFT$_{2}$}}, \href{https://doi.org/10.1007/JHEP08(2023)216}{\emph{JHEP} {\bfseries 08} (2023) 216} [\href{https://arxiv.org/abs/2304.10376}{{\ttfamily 2304.10376}}].

\bibitem{Chen:2023eic}
D.~Chen, X.~Jiang and H.~Yang, \emph{{Holographic TT\textasciimacron{} deformed entanglement entropy in dS3/CFT2}}, \href{https://doi.org/10.1103/PhysRevD.109.026011}{\emph{Phys. Rev. D} {\bfseries 109} (2024) 026011} [\href{https://arxiv.org/abs/2307.04673}{{\ttfamily 2307.04673}}].

\bibitem{He:2023ubi}
P.-Z.~He and H.-Q.~Zhang, \emph{{Holographic timelike entanglement entropy from Rindler method*}}, \href{https://doi.org/10.1088/1674-1137/ad57a8}{\emph{Chin. Phys. C} {\bfseries 48} (2024) 115113} [\href{https://arxiv.org/abs/2307.09803}{{\ttfamily 2307.09803}}].

\bibitem{Narayan:2023zen}
K.~Narayan, \emph{{Further remarks on de Sitter space, extremal surfaces, and time entanglement}}, \href{https://doi.org/10.1103/PhysRevD.109.086009}{\emph{Phys. Rev. D} {\bfseries 109} (2024) 086009} [\href{https://arxiv.org/abs/2310.00320}{{\ttfamily 2310.00320}}].

\bibitem{Shinmyo:2023eci}
K.~Shinmyo, T.~Takayanagi and K.~Tasuki, \emph{{Pseudo entropy under joining local quenches}}, \href{https://doi.org/10.1007/JHEP02(2024)111}{\emph{JHEP} {\bfseries 02} (2024) 111} [\href{https://arxiv.org/abs/2310.12542}{{\ttfamily 2310.12542}}].

\bibitem{Diaz:2023npx}
N.L.~Diaz, J.M.~Matera and R.~Rossignoli, \emph{{Spacetime quantum and classical mechanics with dynamical foliation}}, \href{https://doi.org/10.1103/PhysRevD.109.105008}{\emph{Phys. Rev. D} {\bfseries 109} (2024) 105008} [\href{https://arxiv.org/abs/2311.06486}{{\ttfamily 2311.06486}}].

\bibitem{Kanda:2023jyi}
H.~Kanda, T.~Kawamoto, Y.-k.~Suzuki, T.~Takayanagi, K.~Tasuki and Z.~Wei, \emph{{Entanglement phase transition in holographic pseudo entropy}}, \href{https://doi.org/10.1007/JHEP03(2024)060}{\emph{JHEP} {\bfseries 03} (2024) 060} [\href{https://arxiv.org/abs/2311.13201}{{\ttfamily 2311.13201}}].

\bibitem{Grieninger:2023knz}
S.~Grieninger, K.~Ikeda and D.E.~Kharzeev, \emph{{Temporal entanglement entropy as a probe of renormalization group flow}}, \href{https://doi.org/10.1007/JHEP05(2024)030}{\emph{JHEP} {\bfseries 05} (2024) 030} [\href{https://arxiv.org/abs/2312.08534}{{\ttfamily 2312.08534}}].

\bibitem{Basu:2024bal}
D.~Basu and V.~Raj, \emph{{Reflected entropy and timelike entanglement in TT\textasciimacron{}-deformed CFT2s}}, \href{https://doi.org/10.1103/PhysRevD.110.046009}{\emph{Phys. Rev. D} {\bfseries 110} (2024) 046009} [\href{https://arxiv.org/abs/2402.07253}{{\ttfamily 2402.07253}}].

\bibitem{Anegawa:2024kdj}
T.~Anegawa and K.~Tamaoka, \emph{{Black hole singularity and timelike entanglement}}, \href{https://doi.org/10.1007/JHEP10(2024)182}{\emph{JHEP} {\bfseries 10} (2024) 182} [\href{https://arxiv.org/abs/2406.10968}{{\ttfamily 2406.10968}}].

\bibitem{Wen:2024muv}
Q.~Wen, M.~Xu and H.~Zhong, \emph{{Timelike and gravitational anomalous entanglement from the inner horizon}},  \href{https://arxiv.org/abs/2412.21058}{{\ttfamily 2412.21058}}.

\bibitem{Goswami:2024vfl}
K.~Goswami, K.~Narayan and G.~Yadav, \emph{{No-boundary extremal surfaces in slow-roll inflation and other cosmologies}}, \href{https://doi.org/10.1007/JHEP03(2025)193}{\emph{JHEP} {\bfseries 03} (2025) 193} [\href{https://arxiv.org/abs/2409.14208}{{\ttfamily 2409.14208}}].

\bibitem{Jena:2024tly}
S.S.~Jena and S.~Mahapatra, \emph{{A note on the holographic time-like entanglement entropy in Lifshitz theory}}, \href{https://doi.org/10.1007/JHEP01(2025)055}{\emph{JHEP} {\bfseries 01} (2025) 055} [\href{https://arxiv.org/abs/2410.00384}{{\ttfamily 2410.00384}}].

\bibitem{Xu:2024yvf}
J.~Xu and W.-z.~Guo, \emph{{Imaginary part of timelike entanglement entropy}}, \href{https://doi.org/10.1007/JHEP02(2025)094}{\emph{JHEP} {\bfseries 02} (2025) 094} [\href{https://arxiv.org/abs/2410.22684}{{\ttfamily 2410.22684}}].

\bibitem{Roychowdhury:2025ukl}
D.~Roychowdhury, \emph{{Holographic timelike entanglement and $c$ theorem for supersymmetric QFTs in ($ 0+1 $)d}},  \href{https://arxiv.org/abs/2502.10797}{{\ttfamily 2502.10797}}.

\bibitem{Guo:2025pru}
W.-z.~Guo and J.~Xu, \emph{{A duality of Ryu-Takayanagi surfaces inside and outside the horizon}},  \href{https://arxiv.org/abs/2502.16774}{{\ttfamily 2502.16774}}.

\bibitem{Roychowdhury:2025aye}
D.~Roychowdhury, \emph{{Field theory aspects of $ \eta $-deformed superstring background}},  \href{https://arxiv.org/abs/2503.06294}{{\ttfamily 2503.06294}}.

\bibitem{Giataganas:2025div}
D.~Giataganas, \emph{{Holographic Timelike c-function}}, \href{https://doi.org/10.1103/h7x9-39zs}{\emph{Phys. Rev. D} {\bfseries 114} (2026) 026027} [\href{https://arxiv.org/abs/2505.20459}{{\ttfamily 2505.20459}}].

\bibitem{Diaz:2025aqe}
N.L.~Diaz and R.~Rossignoli, \emph{{Spacetime quantum mechanics for bosonic and fermionic systems}}, \href{https://doi.org/10.1103/zmky-mmwb}{\emph{Phys. Rev. Res.} {\bfseries 8} (2026) 013132} [\href{https://arxiv.org/abs/2506.10250}{{\ttfamily 2506.10250}}].

\bibitem{Roychowdhury:2025ebs}
D.~Roychowdhury, \emph{{Timelike entanglement and central charge for quantum BTZ black holes}}, \href{https://doi.org/10.1016/j.physletb.2025.139846}{\emph{Phys. Lett. B} {\bfseries 869} (2025) 139846} [\href{https://arxiv.org/abs/2507.19813}{{\ttfamily 2507.19813}}].

\bibitem{Heller:2025kvp}
M.P.~Heller, F.~Ori and A.~Serantes, \emph{{Temporal Entanglement from Holographic Entanglement Entropy}}, \href{https://doi.org/10.1103/qlsv-gp22}{\emph{Phys. Rev. X} {\bfseries 15} (2025) 041022} [\href{https://arxiv.org/abs/2507.17847}{{\ttfamily 2507.17847}}].

\bibitem{Nanda:2025tid}
K.K.~Nanda, K.~Narayan, S.~Porey and G.~Yadav, \emph{{dS extremal surfaces, replicas, boundary Renyi entropies in dS/CFT and time entanglement}}, \href{https://doi.org/10.1007/JHEP11(2025)095}{\emph{JHEP} {\bfseries 11} (2025) 095} [\href{https://arxiv.org/abs/2509.02775}{{\ttfamily 2509.02775}}].

\bibitem{Zhou:2025sqq}
Y.~Zhou and P.~Ye, \emph{{Entanglement complexification transition driven by a single non-Hermitian impurity}},  \href{https://arxiv.org/abs/2510.15370}{{\ttfamily 2510.15370}}.

\bibitem{Fujiki:2025rtx}
K.~Fujiki, M.~Kohara, K.~Shinmyo, Y.-k.~Suzuki and T.~Takayanagi, \emph{{Entropic interpretation of Einstein equation in dS/CFT}}, \href{https://doi.org/10.1007/JHEP04(2026)072}{\emph{JHEP} {\bfseries 04} (2026) 072} [\href{https://arxiv.org/abs/2511.07915}{{\ttfamily 2511.07915}}].

\bibitem{Harper:2025lav}
J.~Harper, T.~Kawamoto, R.~Maeda, N.~Nakamura and T.~Takayanagi, \emph{{Non-Hermitian density matrices from timelike entanglement and wormholes}}, \href{https://doi.org/10.1103/j2xw-tcsb}{\emph{Phys. Rev. D} {\bfseries 113} (2026) 126017} [\href{https://arxiv.org/abs/2512.13800}{{\ttfamily 2512.13800}}].

\bibitem{Giataganas:2025ize}
D.~Giataganas, \emph{{Timelike Entanglement Entropy and Renormalization Group Flow Irreversibility}},  \href{https://arxiv.org/abs/2512.16499}{{\ttfamily 2512.16499}}.

\bibitem{Guo:2025dtq}
W.-z.~Guo, \emph{{Spacetime density matrix: formalism and properties}}, \href{https://doi.org/10.1007/JHEP01(2026)128}{\emph{JHEP} {\bfseries 01} (2026) 128} [\href{https://arxiv.org/abs/2508.20397}{{\ttfamily 2508.20397}}].

\bibitem{Liu:2025ctb}
T.~Liu, W.-z.~Guo and S.~He, \emph{{Entanglement of general subregions in time-dependent states}}, \href{https://doi.org/10.1007/JHEP07(2026)087}{\emph{JHEP} {\bfseries 07} (2026) 087} [\href{https://arxiv.org/abs/2512.19955}{{\ttfamily 2512.19955}}].

\bibitem{Li:2026fcr}
Z.-H.~Li and R.-Q.~Yang, \emph{{Black hole interior and time-like entanglement entropy}}, \href{https://doi.org/10.1007/JHEP10(2026)012}{\emph{JHEP} {\bfseries 10} (2026) 012} [\href{https://arxiv.org/abs/2601.18319}{{\ttfamily 2601.18319}}].

\bibitem{Prihadi:2026nua}
H.L.~Prihadi, M.A.R.~Al-Faritsi, R.R.~Firdaus, F.~Khairunnisa, Y.P.~Sarwono and F.P.~Zen, \emph{{Holographic timelike entanglement and subregion complexity with scalar hair}}, \href{https://doi.org/10.1007/JHEP04(2026)174}{\emph{JHEP} {\bfseries 04} (2026) 174} [\href{https://arxiv.org/abs/2601.18310}{{\ttfamily 2601.18310}}].

\bibitem{Narayan:2026wzp}
K.~Narayan, \emph{{de Sitter extremal surfaces, time contours, complexifications and pseudo-entropies}},  \href{https://arxiv.org/abs/2604.00108}{{\ttfamily 2604.00108}}.

\bibitem{Diaz:2026qrf}
N.L.~Diaz, M.~Cerezo and P.~Braccia, \emph{{Unifying spacetime approaches to quantum mechanics}},  \href{https://arxiv.org/abs/2606.12539}{{\ttfamily 2606.12539}}.

\bibitem{Li:2026bof}
Z.-H.~Li and R.-Q.~Yang, \emph{{Linear Growth of Holographic Time-like Entanglement Entropy and Kasner exponents}},  \href{https://arxiv.org/abs/2606.21079}{{\ttfamily 2606.21079}}.

\bibitem{Pal:2026ysc}
J.~Pal and Y.~Shi, \emph{{Holographic Timelike Entanglement and Subregion Complexity in Localized AdS3*S3*T4 Black Holes}},  \href{https://arxiv.org/abs/2607.11641}{{\ttfamily 2607.11641}}.

\bibitem{Bernamonti:2026pxo}
A.~Bernamonti, F.~Galli, M.P.~Heller, F.~Ori and A.~Serantes, \emph{{Temporal Entanglement from Twist Correlators in 2d Conformal Field Theory and Holography}},  \href{https://arxiv.org/abs/2607.14012}{{\ttfamily 2607.14012}}.

\bibitem{Bhattacharyya:2026svh}
A.~Bhattacharyya, M.K.~Hossain, S.~Parihar and S.R.~Roy, \emph{{Holographic timelike complexity for de Sitter}},  \href{https://arxiv.org/abs/2607.29662}{{\ttfamily 2607.29662}}.

\bibitem{Feng:2026grc}
H.~Feng, T.-S.~Chen and S.-F.~Wu, \emph{{Metric Reconstruction from Timelike Entanglement Entropy}},  \href{https://arxiv.org/abs/2608.16517}{{\ttfamily 2608.16517}}.

\bibitem{Ladghami:2026hby}
Y.~Ladghami, F.S.N.~Lobo and T.~Ouali, \emph{{Timelike Entanglement Entropy of Hawking Radiation}},  \href{https://arxiv.org/abs/2602.06833}{{\ttfamily 2602.06833}}.

\bibitem{Saha:2026vin}
A.~Saha, P.~Su{\l}kowski and T.~Takayanagi, \emph{{Pseudo entropy from entanglement entropy}},  \href{https://arxiv.org/abs/2609.14305}{{\ttfamily 2609.14305}}.

\bibitem{Anastasiou:2026bbf}
G.~Anastasiou, I.J.~Araya, A.~Das and J.~Moreno, \emph{{Renormalized pseudoentropy in dS/CFT}}, \href{https://doi.org/10.1007/JHEP05(2026)239}{\emph{JHEP} {\bfseries 05} (2026) 239} [\href{https://arxiv.org/abs/2602.17989}{{\ttfamily 2602.17989}}].

\bibitem{Kanda:2026jyk}
H.~Kanda, T.~Takayanagi and Z.~Wei, \emph{{CFT derivation of entanglement phase transition in pseudo entropy}}, \href{https://doi.org/10.1007/JHEP10(2026)127}{\emph{JHEP} {\bfseries 10} (2026) 127} [\href{https://arxiv.org/abs/2602.22994}{{\ttfamily 2602.22994}}].

\bibitem{Ali-Akbari:2026hys}
M.~Ali-Akbari, \emph{{Holographic Entanglement and Emergent Gravity}},  \href{https://arxiv.org/abs/2608.15862}{{\ttfamily 2608.15862}}.

\bibitem{Ali-Akbari:2026oez}
M.~Ali-Akbari, \emph{{Probe Dependence of the Imaginary Part of HTEE}},  \href{https://arxiv.org/abs/2608.04599}{{\ttfamily 2608.04599}}.

\bibitem{Ori:2026lyk}
F.~Ori and P.-X.~Hao, \emph{{Entanglement entropy in holographic CFTs with generic boundaries}},  \href{https://arxiv.org/abs/2609.21370}{{\ttfamily 2609.21370}}.

\bibitem{Chu:2025sjv}
C.-S.~Chu and H.~Parihar, \emph{{Timelike entanglement entropy with gravitational anomalies}}, \href{https://doi.org/10.1007/JHEP08(2025)038}{\emph{JHEP} {\bfseries 08} (2025) 038} [\href{https://arxiv.org/abs/2504.19694}{{\ttfamily 2504.19694}}].

\bibitem{Alvarez-Gaume:1984zlq}
L.~Alvarez-Gaume and P.H.~Ginsparg, \emph{{The Structure of Gauge and Gravitational Anomalies}}, \href{https://doi.org/10.1016/0003-4916(85)90087-9}{\emph{Annals Phys.} {\bfseries 161} (1985) 423}.

\bibitem{PhysRevD.39.2125}
V.P.~Frolov and K.S.~Thorne, \emph{Renormalized stress-energy tensor near the horizon of a slowly evolving, rotating black hole}, \href{https://doi.org/10.1103/PhysRevD.39.2125}{\emph{Phys. Rev. D} {\bfseries 39} (1989) 2125}.

\bibitem{Tachikawa:2006sz}
Y.~Tachikawa, \emph{{Black hole entropy in the presence of Chern-Simons terms}}, \href{https://doi.org/10.1088/0264-9381/24/3/014}{\emph{Class. Quant. Grav.} {\bfseries 24} (2007) 737} [\href{https://arxiv.org/abs/hep-th/0611141}{{\ttfamily hep-th/0611141}}].

\bibitem{Skenderis:2009nt}
K.~Skenderis, M.~Taylor and B.C.~van Rees, \emph{{Topologically Massive Gravity and the AdS/CFT Correspondence}}, \href{https://doi.org/10.1088/1126-6708/2009/09/045}{\emph{JHEP} {\bfseries 09} (2009) 045} [\href{https://arxiv.org/abs/0906.4926}{{\ttfamily 0906.4926}}].

\bibitem{Brown:1986nw}
J.D.~Brown and M.~Henneaux, \emph{{Central Charges in the Canonical Realization of Asymptotic Symmetries: An Example from Three-Dimensional Gravity}}, \href{https://doi.org/10.1007/BF01211590}{\emph{Commun. Math. Phys.} {\bfseries 104} (1986) 207}.

\bibitem{Hotta:2008yq}
K.~Hotta, Y.~Hyakutake, T.~Kubota and H.~Tanida, \emph{{Brown-Henneaux's Canonical Approach to Topologically Massive Gravity}}, \href{https://doi.org/10.1088/1126-6708/2008/07/066}{\emph{JHEP} {\bfseries 07} (2008) 066} [\href{https://arxiv.org/abs/0805.2005}{{\ttfamily 0805.2005}}].

\bibitem{Compere:2008cv}
G.~Compere and S.~Detournay, \emph{{Semi-classical central charge in topologically massive gravity}}, \href{https://doi.org/10.1088/0264-9381/26/1/012001}{\emph{Class. Quant. Grav.} {\bfseries 26} (2009) 012001} [\href{https://arxiv.org/abs/0808.1911}{{\ttfamily 0808.1911}}].

\bibitem{Keski-Vakkuri:1998gmz}
E.~Keski-Vakkuri, \emph{{Bulk and boundary dynamics in BTZ black holes}}, \href{https://doi.org/10.1103/PhysRevD.59.104001}{\emph{Phys. Rev. D} {\bfseries 59} (1999) 104001} [\href{https://arxiv.org/abs/hep-th/9808037}{{\ttfamily hep-th/9808037}}].

\bibitem{Deser:1991qk}
S.~Deser and X.~Xiang, \emph{{Canonical formulations of full nonlinear topologically massive gravity}}, \href{https://doi.org/10.1016/0370-2693(91)91704-Y}{\emph{Phys. Lett. B} {\bfseries 263} (1991) 39}.

\bibitem{Carlip:2008qh}
S.~Carlip, \emph{{The Constraint Algebra of Topologically Massive AdS Gravity}}, \href{https://doi.org/10.1088/1126-6708/2008/10/078}{\emph{JHEP} {\bfseries 10} (2008) 078} [\href{https://arxiv.org/abs/0807.4152}{{\ttfamily 0807.4152}}].

\bibitem{Chen:2025xlo}
C.Y.-R.~Chen, E.~Joung, K.~Mkrtchyan and J.~Yoon, \emph{{Higher-order chiral scalar from boundary reduction of 3d higher-spin gravity}}, \href{https://doi.org/10.1007/JHEP05(2025)186}{\emph{JHEP} {\bfseries 05} (2025) 186} [\href{https://arxiv.org/abs/2501.16463}{{\ttfamily 2501.16463}}].

\bibitem{Chen:2011yx}
B.~Chen and J.~Long, \emph{{High Spin Topologically Massive Gravity}}, \href{https://doi.org/10.1007/JHEP12(2011)114}{\emph{JHEP} {\bfseries 12} (2011) 114} [\href{https://arxiv.org/abs/1110.5113}{{\ttfamily 1110.5113}}].

\end{thebibliography}\endgroup
  
\end{document}